\documentclass[titlepage]{article}

\usepackage{graphicx}

\usepackage[]{natbib}

\usepackage{xspace}
\usepackage{amsmath,amssymb} 
\usepackage[margin=1.5in]{geometry}

\usepackage{textcomp} 
\usepackage[colorlinks=true]{hyperref}

\newcommand{\prebot}{preB\"otC\xspace}
\newcommand{\botc}{B\"otC\xspace}
\newcommand{\pico}{PiCo\xspace}

\newcommand{\ER}{Erd\H{o}s-R\'enyi\xspace}
\newcommand{\kavg}{k_\textrm{avg}}
\newcommand{\pI}{p_I}
\newcommand{\gE}{g_E}
\newcommand{\gI}{g_I}
\newcommand{\gleak}{g_L}
\newcommand{\ENa}{E_\textrm{Na}}
\newcommand{\EK}{E_\textrm{K}}
\newcommand{\Eleak}{E_L}
\newcommand{\Em}{\theta_m}
\newcommand{\En}{\theta_n}
\newcommand{\Emp}{\theta_{m,p}}
\newcommand{\Eh}{\theta_h}
\newcommand{\sm}{\sigma_m}
\newcommand{\sn}{\sigma_n}
\newcommand{\smp}{\sigma_{m,p}}
\newcommand{\sh}{\sigma_h}
\newcommand{\taunb}{\bar{\tau}_{n}}
\newcommand{\tauhb}{\bar{\tau}_h}
\newcommand{\gK}{g_\textrm{K}}
\newcommand{\gNa}{g_\textrm{Na}}
\newcommand{\gNaP}{g_\textrm{Na,p}}
\newcommand{\EsynE}{E_\textrm{syn,E}}
\newcommand{\EsynI}{E_\textrm{syn,I}}
\newcommand{\Esyn}{\theta_\textrm{syn}}

\newcommand{\ssyn}{\sigma_\textrm{syn}}
\newcommand{\tausyn}{\bar{\tau}_\textrm{syn}}
\newcommand{\Iapp}{I_\textrm{app}}
\newcommand{\excnodes}{\mathcal{N}_E}
\newcommand{\inhnodes}{\mathcal{N}_I}
\newcommand{\Ileak}{I_L}
\newcommand{\INa}{I_\textrm{Na}}
\newcommand{\IK}{I_\textrm{K}}
\newcommand{\INaP}{I_\text{Na,p}}
\newcommand{\Isyn}{I_\text{syn}}
\newcommand{\xbin}{\mathbf{x}^\mathrm{bin}}

\newcommand{\xbtai}[1]{x_{#1}^\mathrm{BTA}}
\newcommand{\nbursts}{n_\mathrm{bursts}}
\newcommand{\xsmooth}{\mathbf{x}^\mathrm{filt}}
\newcommand{\xsmoothi}[1]{x_i^\mathrm{filt}}
\newcommand{\popint}{x^\mathrm{int}}
\newcommand{\xavg}{\bar{x}}
\newcommand{\sync}{\chi}
\newcommand{\xsmoothavg}{\bar{x}^\mathrm{filt}}

\newcommand{\dee}{\mathrm{d}}
\newcommand{\kintra}{k_{\rm intra}}
\newcommand{\kinter}{k_{\rm inter}}
\newcommand{\tj}{t^{i*}_j}
\newcommand{\tjnext}{t^{i*}_{j+1}}
\newcommand{\tl}{t^{\bar{i}*}_\ell}
\newcommand{\zavg}{\zeta_{\rm avg}}
\newcommand{\popone}{Pop.\ 1\xspace}
\newcommand{\poptwo}{Pop.\ 2\xspace}
\newcommand{\si}{}
\newcommand{\micro}{\textmu}

\begin{document}
\title{
Different roles for inhibition in the rhythm-generating respiratory network
}

\author{Kameron Decker Harris \and 
  Tatiana Dashevskiy \and
  Joshua Mendoza \and
  Alfredo J.\ Garcia III \and
  Jan-Marino Ramirez \and 
  Eric Shea-Brown}
  
\date{\today}

\maketitle

\begin{abstract}
 Unraveling the interplay of excitation and inhibition within
 rhythm-generating networks remains a fundamental issue in neuroscience.
We use a biophysical model to investigate the different roles 
of local and long-range inhibition in 
the respiratory network, 
a key component of which is the pre-B\"otzinger complex inspiratory microcircuit.
Increasing inhibition within the microcircuit
results in a limited number of out-of-phase neurons 
before rhythmicity and synchrony degenerate.
Thus, unstructured local inhibition is destabilizing and
cannot support the generation of more than one rhythm.  
A two-phase rhythm requires restructuring the network into two 
microcircuits coupled by long-range inhibition in the manner of a half-center.
In this context, inhibition leads to greater stability of the two out-of-phase rhythms.
We support our computational results with in vitro recordings from 
mouse pre-B\"otzinger complex.
Partial excitation block leads to increased rhythmic variability,
but this recovers following blockade of inhibition.
Our results support the idea that local inhibition in the pre-B\"otzinger complex
is present to allow for descending control of synchrony or
robustness to adverse conditions like hypoxia.
We conclude that the balance of inhibition and excitation
determines the stability of rhythmogenesis, 
but with opposite roles within and between areas.
These different inhibitory roles may apply to a variety of rhythmic behaviors
that emerge in widespread pattern generating circuits of the nervous system.
\end{abstract}

\section*{New \& Noteworthy}

The roles of inhibition within the pre-B\"otzinger complex (\prebot) 
are a matter of debate.
Using a combination of modeling and experiment,
we demonstrate that inhibition affects synchrony, 
period variability,
and overall frequency of the \prebot
and coupled rhythmogenic networks.
This work expands our understanding of ubiquitous motor and cognitive
oscillatory networks.

\section{Introduction}

Rhythmic activity is critical for the generation of behaviors such as 
locomotion and respiration, as well as apparently 
non-rhythmic behaviors including olfaction, 
information processing, encoding, learning and memory 
\citep{marder2001,buzsaki2006,kopell2010,ainsworth2012,skinner2012,missaghi2016}.
These rhythms arise from central pattern generators (CPGs), 
neuronal networks located within the central nervous system
that are capable of generating periodic behavior
due to their synaptic and intrinsic membrane properties
\citep{marder2001,grillner2006,grillner2009,kiehn2011}.

An increasingly important concept is that a given behavior
may involve the interaction between several rhythmogenic microcircuits
\citep{anderson2016, ramirez2016}.
In the neocortex, multiple rhythms and mechanisms are involved in a 
variety of cortical processes \citep{buzsaki2006}. 
In breathing, which consists of the three dominant respiratory phases---inspiration, 
post-inspiration, and expiration---each phase seems to be generated by its own autonomous,
excitatory microcircuit,
sub-populations of the overall network which act as rhythm-generating modules 
\citep{anderson2016,lindsey2012}.
The timing between these excitatory microcircuits is established by inhibitory interactions.
In locomotion, each side of the spinal cord contains rhythmogenic microcircuits 
that are similarly coordinated 
by inhibitory mechanisms in order to establish left-right alternation
\citep[e.g.\ ][]{kiehn2011}.
Assembling a behavior by combining different microcircuits may imbue a network with 
increased flexibility. 
This strategy could also facilitate the integration and synchronization
of one rhythmic behavior with another.
Sniffing, olfaction, whisking, and rhythmic activities in hippocampus and locus coeruleus
are all rhythmically coupled to the inspiratory rhythm 
generated in the pre-B\"{o}tzinger complex
(\prebot) \citep{sara2009, moore2013, ferguson2015,ramirez2016, huh2016}.
This small microcircuit, located in the ventrolateral medulla, 
is the essential locus for the generation of breathing 
\citep{smith1991,tan2008,gray2001,schwarzacher2011,ramirez1998}.

First discovered a quarter of a century ago, 
the \prebot is among the best understood microcircuits \citep{smith1991}. 
It continues to generate fictive respiratory rhythm activity when isolated in vitro,
reliant on excitatory neurotransmission.
Rhythmicity in the \prebot ceases when glutamatergic synaptic mechanisms are blocked,
while it persists following the blockade of synaptic inhibition.
However, almost 50\% of the \prebot neurons are inhibitory
\citep{shao1997,winter2009,morgado-valle2010,hayes2012}.
Despite the abundance of inhibitory neurons, 
the majority of neurons in the \prebot are rhythmically active in phase with inspiration.
A small group of approximately 9\% of neurons in the \prebot
are inhibited during inspiration and discharge in phase with expiration
\citep{morgado-valle2010,nieto-posadas2014, carroll2013}.
A recent optogenetic study by \citet{sherman2015}
showed that stimulation of glycinergic inhibitory \prebot neurons can
delay or halt a breath, 
and inhibition of those neurons can increase the magnitude of a breath.
This is consistent with pharmacological agonist-antagonist 
experiments by \citet{janczewski2013} which found that inhibition
can modulate rhythm frequency or trigger apnea but is not essential
for rhythm generation.
The inhibitory population may thus be an ``actuator'' 
that allows descending pathways to control respiration.
However, with only a few studies available,
the role of these inhibitory \prebot neurons is not well-understood.

These experimental findings raise important questions:
What is the role of inhibitory neurons within this microcircuit 
\citep{cui2016}?
Why does the \prebot generate primarily one rhythmic phase
despite the presence of numerous inhibitory neurons? 
Our modeling study arrives at the conclusion that 
this microcircuit can only generate one rhythmic phase.
Synaptic inhibition seems to primarily serve to 
titrate the strength of this single rhythm 
while creating a small number of apparently anomalous expiratory cells.
In order to generate more than one phase,
it is necessary to assemble a  network where excitatory microcircuits 
are segmented, via inhibition, into different compartments.
Mutually-inhibitory circuits have been proposed for the 
inspiration--active expiration network
\citep{smith2013,molkov2013,koizumi2013,onimaru2015}
and \prebot--post-inspiratory complex (\pico) networks \citep{anderson2016}.

The novelty of our theoretical study lies in two conceptually important findings:
A single microcircuit is unable to generate more than one phase based on the currently
known network structure, and 
the generation of different phases necessitates the
inhibitory interaction between excitatory microcircuits.
Based on these findings we propose that the generation of rhythm and phase arise from
separate network-driven processes. 
In these two processes, inhibition plays fundamentally different roles: 
local inhibition promotes desynchronization within a microcircuit, 
while long-range inhibition establishes phase relationships between microcircuits.
Consistent with our proposal is the observation that breathing does not depend
on the presence of all three phases at any given time.
In gasping and some reduced preparations, 
the respiratory network generates a one-phase rhythm consisting of inspiration only.
Under resting conditions, breathing primarily oscillates between inspiration and post-inspiration.
This eupneic rhythm also involves a late expiratory phase according to \citet{richter2014}.
Under high metabolic demand or coughing, 
another phase is recruited in form of active expiration.
This modular organization may be a fundamental property of rhythm generating networks.

\section{Materials and Methods}
\label{sec:materialsandmethods}
\subsection{\prebot network simulations}

\label{sec:model}

We model the \prebot network as a simple directed
\ER random graph on $N=300$ nodes,
where edges are added at random with fixed probability.
We denote a directed edge from node $j$ to node $i$ as $j\to i$.
The connection probability
$p = (\kavg/2)/(N-1)$
so that the expected total degree,
that is the in-degree plus the out-degree,
of a node is $\kavg$, 
which we vary.
We prefer to parametrize these networks by degree
$\kavg$ 
rather than 
$p$,
since in this case our results do not depend
on $N$ once it is large 
\citep{bollobas1998}.

Each node is of type
bursting (B), tonic spiking (TS), or quiescent (Q)
with corresponding probabilities 
25\%, 45\%, and 30\%
\citep{pena2004,delnegro2005}.
Neurons are inhibitory with probability $\pI$,
another parameter,
and all projections from an inhibitory neuron are inhibitory.
The sets of excitatory and inhibitory nodes are denoted
$\excnodes$ and $\inhnodes$.
Edges are assigned a maximal conductivity
$\gE$ for excitatory connections and
$\gI$ for inhibitory connections.
In our parameter sweeps,
we vary these conductivities 
over the range 2--5 nS.
This matches the postsynaptic potential deflections
observed in experiments 
(typical IPSPs: -1.2 to -1.8 mV, 
EPSPs: 1.6 to 2.3 mV;
data from Aguan Wei).

We use ``model 1'' from 
\citet{butera1999}
as the dynamical equations for
bursting, tonic spiking, and quiescent neurons.
All parameters, given in Table~\ref{tab:param},
are shared among the dynamical types
with the exception of the leak conductance
$\gleak$ which is adjusted for the desired dynamics
(B, TS, Q).
Parameter values besides $\gleak$ are taken from 
\cite{park2013}, 
most of which are the same or close to the original values
chosen by \citet{butera1999}.
With the chosen parameters, 
the bursting neurons fire 6-spike bursts
every 2.4 s,
and the tonic spikers fire
3.5 spikes per second.

\begin{table}[t]
  \begin{minipage}[c]{0.38\linewidth}
    \begin{center}
      \begin{tabular}{c|c}
        Parameter & Value\\
        \hline
        \hline
        $C$ & 21 pF \\
        $\ENa$  & 50 mV \\
        $\EK$ & -85 mV \\
        $\Eleak$ & -58 mV \\
        $\Em$ & -34 mV\\
        $\En$ & -29 mV \\
        $\Emp$ & -40 mV \\
        $\Eh$ & -48 mV\\
        $\sm$ & -5 mV\\
        $\sn$ & -4 mV \\
        $\smp$ & -6 mV\\
        $\sh$ & 5 mV\\
        $\taunb$ & 10 ms \\
        $\tauhb$ & 10,000 ms \\
        $\gK$ & 11.2 nS \\
        $\gNa$ & 28 nS \\
        $\gNaP$ & 1 nS \\
        $\Iapp$ & 0 pA \\
        $\gleak^{\rm (B)}$ & 1.0 nS \\
        $\gleak^{\rm (TS)}$ & 0.8 nS \\
        $\gleak^{\rm (Q)}$ & 1.285 nS \\
        \hline
        $\EsynE$ & 0 mV \\
        $\EsynI$ & -70 mV\\
        $\Esyn$ & 0 mV \\ 
        $\ssyn$ & -3 mV \\
        $\tausyn$ & 15 ms
      \end{tabular}
    \end{center}
  \end{minipage} \hfill
  \begin{minipage}[c]{0.55\linewidth}
    \begin{center}
      \caption{Parameters for the network model
        are taken from the literature \citep{butera1999,park2013}.
        We modify $\gleak$ for quiescent (Q), tonic spiking (TS),
        and intrinsically bursting (B) cells.
        The system of equations is simulated in the given units,
        so that no conversions are necessary.
        Those parameters below the lower horizontal break 
        are for the synaptic dynamics.
      }
      \label{tab:param}
    \end{center}
  \end{minipage}
\end{table}

The full system of equations is 
\begin{gather}
  \label{eq:odes}
  \begin{aligned}
    \dot{V} &= - \left( \Ileak + \INa + \IK + \INaP + \Isyn -\Iapp
    \right) / C\\
    \dot{h} &= \left( h_\infty (V) - h \right) / \tau_h (V) \\
    \dot{n} &= \left( n_\infty (V) - n \right) / \tau_n (V)
  \end{aligned}
\end{gather}
with currents calculated as
\[
\begin{aligned}
\Ileak &= \gleak (V - E_\text{L}) \\
\INa &= \gNa m_\infty^3(V) (1 - n) (V - \ENa) \\
\IK &= \gK n^4 (V - \EK) \\
\INaP &= \gNaP m_{\text{p},\infty}(V) h (V - \ENa) ,
\end{aligned}
\]
and the activation and time constants are
\begin{align*}
x_\infty(V) &=   \frac{1}{1 + \exp \left( (V - \theta_x)/\sigma_x \right)} \\
\tau_x(V) &=  \frac{\bar{\tau}_x}{
  \cosh \left((V - \theta_x)/(2 \sigma_x) \right)}.
\end{align*}

To model network interactions, 
we model synaptic dynamics with first-order kinetics
\citep{destexhe1994}.
The synaptic current neuron $i$ receives is
\[
I_{\text{syn},i} = 
 \sum_{j \in \excnodes: j \to i}
 \gE s_{ij} \left(V_i - \EsynE \right)
 +  
 \sum_{j \in \inhnodes: j \to i}
 \gI s_{ij} \left(V_i - \EsynI \right),
\]
where 
$\gE$ and $\gI$
are the maximal excitatory and inhibitory synapse conductances.
The reversal potentials 
$\EsynE$ and $\EsynI$
for excitatory and inhibitory synapses,
shown in Table~\ref{tab:param},
correspond the appropriate values for
glutamatergic and glycinergic or GABAergic synapses.
The variables
$s_{ij}$ 
represent the open fraction of channels between cells $j$ and $i$, 
and they are governed by the differential equations
\[
\begin{aligned}
\dot{s}_{ij} &= 
\left( (1 - s_{ij}) m^{(ij)}_\infty(V_j) - s_{ij} \right) / 
\tau_{\text{syn}} \\
m^{(ij)}_\infty (V_j) &= 
\frac{1}{1 + \exp \left( (V_j - \Esyn ) 
/ \sigma_\text{syn} \right)} .
\end{aligned}
\]
Excitatory and inhibitory synapses share the parameters
$\tausyn$, $\Esyn$, and $\ssyn$ (Table~\ref{tab:param}).

Each model run starts from random initial conditions and lasts
100 s of simulation time with 1 ms time resolution.
The first 20 s of transient dynamics are removed before postprocessing.
Rather than save all state variables during long runs,
we record a binary variable 
for each neuron that indicates whether or not the neuron fires a spike
in the given time step.
A spike is registered when
$V$ surpasses -15 mV for the first time
in the previous 6 ms.
This spike raster is then stored as a sparse matrix.
The simulation code is configurable to output voltage traces or all state variables;
these were examined during development to check that the model and
spike detection function correctly.

We examine the effects of network connectivity, inhibition, and 
synaptic strength on the dynamics of our model by varying
$\kavg$, $\pI$, $\gE$, and $\gI$.
To capture the interactions of these parameters,
we sweep through
all combinations of parameters in the ranges
$\kavg = 1.0, 1.5, \ldots, 12.0$;
$\pI = 0.00, 0.05, \ldots, 1.00$;
$\gE = 2, 3, \ldots, 5$ nS;
and
$\gI = 2, 3, \ldots, 5$ nS,
with 8 repetitions of each combination.
The only randomness in the model is randomness
present in the graphs and initial conditions, 
since the dynamics are deterministic.
This amounts to 61,824 graph generation, simulation, and postprocessing steps.
Network generation, simulations,
and postprocessing were performed with custom
code available from the
first author at \url{http://github.com/kharris/prebotc-graph-model}.
The code was written in Python and C++, and some analysis was performed with MATLAB.
Numerical integration used backwards differentiation formulae
in VODE called via {\tt scipy.integrate.ode},
suitable for stiff equation systems.
We experimented with the tolerance to be sure it resolves all timescales.
We used the Hyak cluster at the University of Washington to conduct parameter sweeps.
Each simulated 100 s took less than 3 hours and could be performed on a standard consumer machine.

\subsection{Two population network model}
\label{sec:twopop}

The \prebot is thought to be connected to another microcircuit,
alternately the \botc, \pico, and lateral parafacial group,
in a mutually inhibitory manner 
\citep{smith2013,molkov2013,huckstepp2016,anderson2016}
which allows them to generate stable two-phase rhythms
as in a half-center oscillator \citep{marder2001}. 
We study this case with a two microcircuit model, 
a  where each microcircuit 
is represented by a different population of cells (\popone and \poptwo);
 we arbitrarily refer to the \prebot as \popone.

We use a two group stochastic block model for the network.
The stochastic block model \citep{holland1983} 
is a generalization of the directed \ER random graph,
where the connection probability varies depending on the population label of each neuron.
Each population has recurrent connections from excitatory to all other cells, 
with each connection occurring with a fixed probability. 
As we describe below,
we vary probabilities of connections from inhibitory neurons to other neurons in the same population (intra-group) and in the other population (inter-group).

Let $N_1$ be the number of neurons in \popone and 
$N_2$ be the number of neurons in \poptwo. 
We assume $N_1 = N_2 = 300$, so the network has a total of 600 neurons.  
To generate this network we begin by assigning each neuron to one of the two populations. 
We then assign each neuron a type: quiescent, tonic or bursting, using the same method as the single population model. 
Afterwards, we randomly assign neurons to be inhibitory with probability $\pI = 0.5$ \citep{shao1997,winter2009,hayes2012,morgado-valle2010};
otherwise they are excitatory. 
We then assign connections to the neurons with probabilities: 
\[
P^{(I)} = \begin{bmatrix} \frac{\kintra}{N_1-1}&\frac{\kinter}{N_2}\\ 
                      \frac{\kinter}{N_1}& \frac{\kintra}{N_2-1} \end{bmatrix}, 
                      \quad
P^{(E)} = \begin{bmatrix} \frac{3}{N_1-1}& 0\\ 
                      0& \frac{3}{N_2-1} \end{bmatrix},\;
\]
where $0 \leq \kintra , \kinter \leq 4$.
The matrix entries $(i,j)$ are the probability of a connection between 
an inhibitory or excitatory
neuron in population $i$
to a neuron in population $j$.
This model allows us to tune between a half-center 
network containing only inter-group inhibition and 
a network with equal amounts of both intra- and inter-group inhibition.

The matrix $P^{(E)}$ contains the probability of
connection for
a projecting excitatory neuron.
It is diagonal, reflecting the assumption that excitatory neurons only
project within the local population,
and each excitatory neuron has an average out-degree of 3.
The matrix $P^{(I)}$
describes the probability of connection for inhibitory projecting neurons. 
The variable $\kintra$ is the expected
number of projections per inhibitory neuron
to other neurons within its own population,
and $\kinter$
is the expected number of projections from an inhibitory neuron to 
neurons in the other population. 
We normalize these values in the matrix to ensure that the 
average in-degree is the sum of the columns and and out-degree is the sum of the rows,
both equal to $\kintra + \kinter + 3$.
The total inhibitory degrees then depend on the values of $\kintra$ and $\kinter$, 
which affect only the inhibitory connection probabilities.
Unless explicitly stated, connections are assigned a fixed conductance of $\gE = \gI = 2.5$ nS for excitatory
and inhibitory connections. 

We examine the effects of inhibition both within a population and between populations. 
To do this, we sweep through the parameters $\kintra, \kinter = 0.0, 0.5, \ldots, 4.0$
and simulate 8 realizations 
(i.e., samples from the distribution of random graphs with these parameters) 
for each parameter pair.
This leads to 648 graph generation,
simulation, 
and post processing steps. 
As for the single population model, all code is available at 
\url{http://github.com/kharris/prebotc-graph-model} .

\subsection{Slice experiments}
\label{sec:experiments}

Brainstem transverse slices were prepared from CD1 mice
(P7--12). All experiments were performed with the approval of the 
Institute of Animal Care and Use Committee of the Seattle Children's Research Institute. 
Mice were maintained with rodent diet and water available ad libitum 
in a vivarium with a 12 h light/dark cycle at 22$^{\circ}$C.  
Thickness of slices containing the \prebot varied between
550-650 \textmu m. 
Slices were placed into the recording chamber with
circulating artificial cerebrospinal fluid (aCSF) containing 
NaCl 118 mM, KCl 3 mM, 
$\mathrm{CaCl_2}$ 1.5 mM, MgCl 1 mM, 
$\mathrm{NaHCO_3}$ 25 mM, 
$\mathrm{NaH_2 PO_4}$ 1 mM,
d-glucose 30 mM and equilibrated with 
95\% $\mathrm{O_2}$ and 5\% $\mathrm{CO_2}$, 
pH 7.4. 
We maintained the 
temperature of the bath at 31$^{\circ}$C, 
with an aCSF circulation rate of 15 mL/min. 
Rhythmic activity of \prebot was induced by slow
up-regulation of KCl concentration from 3 mM to 8 mM in aCSF.  
The details of the technique are described in \citet{ramirez1997}
and \citet{anderson2016}.

We recorded
extracellular neuronal
population activity in the \prebot region with a protocol 
that first measured the control activity,
then activity following application
of a partial excitation block,
and finally with an additional complete block of inhibition.
We used 700~nM DNQX disodium salt, 
a selective non-NMDA receptor antagonist
which blocks glutamatergic ion channels generating
fast excitatory synaptic inputs, 
to effect the partial excitation block.
Picrotoxin
(PTX), an ionotropic $\mathrm{GABA_A}$
receptor antagonist
blocking inhibitory chloride-selective channels,
was used at 20 or 50~\textmu M to shut down inhibition.
Both concentrations of PTX were equally effective at blocking inhibition.
DNQX disodium salt and PTX were obtained from
Sigma-Aldrich, St.\ Louis, MO.
After application of either drug,
we waited 5 min for the drugs to take effect and used at least
10 min of data to measure the resulting  rhythm. 

In additional experiments,
we supplemented the extracellular population-level 
data with multi-electrode recordings in the contralateral \prebot.
Extracellular neural activity from the transverse 
medullary slice was recorded on a 16 channel commercial 
linear multi-array electrode (model: Brain Slice Probe, Plexon, Dallas, TX).  
Each electrode had a recording surface of 15 microns and 
interelectrode spacing was fixed at 50 microns.  
Neural signals were amplified and recorded using the Omni-Plex D system (Plexon).   
Wide-band data was filtered with a Butterworth lowpass filter, 200 Hz cutoff,
and spike sorting was performed offline and post-hoc using Offline Sorter v4.1.0 (Plexon).
Specifically, individual unit waveforms were detected and sorted 
using principle component analysis, visualized in a three-dimensional cluster view.
Waveforms were detected and sorted using Offline Sorter
with manual cluster cutting single electrode-based feature spaces. 
Care was taken to follow nonstationarities in waveform shapes 
in assigning spikes to separate units, 
and auto- and cross-correlation histograms were examined as a check on sorting results
\citep{lewicki1998}.
All neurons with good isolation were kept for analysis.

We kept only those slices that initially showed robust rhythms,
as determined by the experimentalist.
We performed a total of 5 multi-electrode experiments and discarded
one in which the rhythm went away after application of DNQX 
and never recovered.
We recorded extracellularly from 15 slices and excluded 2 outliers from 
statistical analysis,
because their rhythms slowed considerably more than the others with DNQX.
In vitro slice data were analyzed by hand using Axon pClamp 
(Molecular Devices, Sunnyvale, CA)
to extract burst locations and amplitudes, which were exported to a table
for analysis using custom Python programs available at
\url{http://github.com/kharris/prebotc-graph-model}.

\subsection{Postprocessing}
\label{sec:postprocessing}

Because of the large number of simulations needed to explore the parameter
space, we can examine only a small fraction of the simulations by eye and
must rely on summary statistics to characterize the dynamics.

\subsubsection{Binning and filtering}

\label{sec:filtering}

First, the spike raster data is aggregated into 
50 ms
bins of spike counts
to compress the size of the matrix.
We denote the spike raster vector timeseries $\xbin(t)$.
The unbinned spike rasters are then convolved with
a Gaussian kernel
$k(t) = \left( \sigma \sqrt{2 \pi} \right)^{-1}
\exp \left( - \frac{1}{2} t^2/\sigma^2 \right)$,
where
$\sigma = 60 $ ms,
to produce the continuous timeseries
$\xsmooth(t) = (k * \mathbf{x})(t)$,
which is then downsampled to the same time bins.
To characterize the overall population output,
we compute what we call the integrated trace $\popint(t)$.
This is defined as the lowpass-filtered
population average, 
where
the population average
$\xavg (t) = \frac{1}{N}\sum_{i=1}^N x_{i}(t)$.
We use a second-order Butterworth filter with
cutoff frequency 4 Hz.
The integrated trace is normalized to have units of 
spikes per second per neuron.

\subsubsection{Synchrony statistic}
\label{sec:synchrony}

Our principle aim is to quantify how different networks give rise
to varying degrees of synchrony across the population of bursting 
neurons.
We choose to characterize the overall synchrony of the population
with one statistic  \citep{golomb2007, masuda2004}
\begin{equation}
  \label{eq:chi}
  \sync = \left(
    \frac{\langle \xsmoothavg(t)^2 \rangle_t
      - \langle \xsmoothavg(t) \rangle_t^2}{
      \frac{1}{N}\sum_{i=1}^N \left[ \langle \xsmoothi{i}(t)^2 \rangle_t
      - \langle \xsmoothi{i}(t) \rangle^2_t \right] }
\right)^{1/2}
\end{equation}
where the angle brackets
$\langle \cdot \rangle_t$ 
denote averaging over the timeseries
and 
$\xsmoothavg(t) = \frac{1}{N}\sum_{i=1}^N \xsmoothi{i}(t)$.
The value of $\sync$ is between 0 and 1.
With perfect synchrony,
$\xsmoothi{i}(t) = \xsmoothavg(t)$ for all $i$,
then we will find $\sync = 1$.
With uncorrelated signals $\xsmoothi{i}(t)$,
then $\sync = 0$.
Examples of network activity for different values of 
$\sync$ are shown in Fig.~\ref{fig:rasters}.

\subsubsection{Burst detection and phase analysis}
\label{sec:phases}

The respiratory rhythm is generated by synchronized bursts of
activity in the \prebot.
In order to identify these bursts in the integrated traces,
we needed a method of peak-detection that identifies large bursts
but ignores smaller fluctuations.
To do this we identify times 
$t^*$
in the integrated timeseries 
$\popint(t)$,
where 
$\popint(t^*)$
is an absolute maxima over a window of size 600 ms 
(12 time bins to either side of the identified maximum),
and its value is above the 75th percentile of the full 
integrated timeseries.
This ensures that the detected bursts are large-amplitude, 
reliable maxima of the timeseries.

Using the detected burst peak times  
$t^*_1, t^*_2, \ldots, t^*_{\nbursts}$, 
we can examine the activity of individual neurons triggered 
on those events, the burst triggered average (BTA).
The time between consecutive bursts is irregular, 
so in order to compute averages over many events, 
we rescale time into a uniform phase variable 
$\phi \in [-\pi, \pi]$.
A phase
$\phi=0$ 
happens at the population burst, 
while
$\phi=-\pi \equiv \pi \pmod{2\pi}$ 
occurs in-between bursts.
To define this phase variable,
we rescale the half-interval 
$\left[(t^*_n-t^*_{n-1})/2, t^*_n \right]$
preceding burst $n$ to 
$[-\pi, 0]$.
Similarly, we rescale the other half-interval
$\left[ t^*_n, (t^*_{n+1}-t^*_n)/2 \right]$ 
which follows burst
$n$
to 
$[0, \pi]$.
Each rescaling is done using linear interpolation of the
binned spike rasters.
Let $\Phi(t)$ denote the mapping from time $t$ to the
phase.
Then the BTA
activity of neuron $i$ is
\begin{equation}
\xbtai{i}(\phi) = 
\frac{1}{\nbursts} \sum_{j=1}^{\nbursts}
\int_{-(t^*_j-t^*_{j-1})/2}^{(t^*_{j+1}-t^*_j)/2} 
\xsmoothi{i} \left(t^*_j+t\right) \, 
\delta \left(\Phi\left(t^*_j+t\right)-\phi \right) \dee t,
\label{eq:bta}
\end{equation}
where $\delta(\cdot)$ is the Dirac delta measure which ensures that 
$\xsmoothi{i}$ is sampled at the correct phase.

The BTAs exhibit two characteristic shapes.
The first shape is peaked at a particular value of $\phi$;
these neurons are phasic bursters.
Of course, most phasic bursters take part in the overall population
rhythm and have their BTA maximum near zero. 
Cells that are in-phase with the population rhythm are
{\it inspiratory}.
However, there are some bursters with a BTA peak near $\pi$,
and we call these out-of-phase cells {\it expiratory}.
The second shape is weakly peaked or flat;
these neurons are {\it tonic}.

We define a complex-valued
{\it phase-locking variable}
$z_i$
as the  circular average of the BTA
normalized by its integral:
\begin{equation}
  \label{eq:OP}
  z_i = 
  \frac{
    \int_{-\pi}^{\pi} \xbtai{i}(\phi) e^{\mathrm{i} \phi} \dee \phi
  }
  {
    \int_{-\pi}^{\pi} \xbtai{i}(\phi) \dee \phi
  }.
\end{equation}
Normalization allows us to compare cells 
with different firing rates.
The magnitude of phase-locking
(peakedness of $\xbtai{i}$)
is quantified by the magnitude $|z_i|$.
We use the argument $\arg(z_i)$ to define the dominant phase
of a cell's activity. 
These phase-locking variables are similar to the order parameters
used to study synchrony \citep{arenas2008}.
We classify cell $i$ as inspiratory, expiratory, tonic, or silent by:
\begin{enumerate}
\item Silent: firing rate is less than 0.1 Hz,
\item Inspiratory: $|z_i|> 0.2$ and $|\arg(z_i)| \leq \pi/2$,
\item Expiratory:  $|z_i|> 0.2$ and $|\arg(z_i)| > \pi/2$,
\item Tonic: otherwise.
\end{enumerate}

\subsubsection{Two population phase analysis}
\label{sec:phases_2}

For the two microcircuit model,
we are also interested in the phase relationship between the two populations. 
To study this, we examine the burst-by-burst phase differences between the two populations'
integrated traces and extract descriptive statistics of the phase differences.
The $N_1$ neurons in \popone and $N_2$ neurons in \poptwo 
define two separate groups that we analyze as in Sections~\ref{sec:filtering}, \ref{sec:synchrony},
and \ref{sec:phases}.
Note that because of the symmetry of $P^{(E)}$ and $P^{(I)}$, 
\popone and \poptwo are statistically equivalent.
The burst times define two vectors 
$t^{1*}$ and $t^{2*}$,
where $t^{i*}_j$ is the time for the $j\rm th$ peak in the signal of population $i=1$ or 2. 
\popone is set as the reference signal for phase analysis.
We then define a window with respect to the reference as 
$W_j = [\tj,\tjnext]$, where $i$ is the chosen reference signal. 
For each peak $\ell$ in the non-reference signal, 
which we write as $\tl$, 
we find the reference window $W_j$ so that $\tl \in W_j$. 
In other words, for each peak in
the non-reference signal we find the two peaks it lies between in the reference
signal; we say that these peaks delineate the reference window. 
Once we have the reference window to use for the given peak, 
we define the phase difference between the two signals as 
$\theta_i = \frac{\tjnext - \tl}{\tjnext-\tj} \in [0,1]$.

For an accurate description of the overall phase
difference between the signals, 
we use directional statistics
\citep{jammalamadaka2001},
which account for the fact that $\theta = 0$ and
$1$ are identified. We can imagine that each phase
difference is mapped to a circle, where we can then
calculate the average position of those phase
differences and how spread out the values are on that circle
with respect to that average. 
To do this, we map the $\theta_i$ onto the unit circle using
the equation 
$\zeta_k = e^{2 \pi \mathrm{i} \theta_k}$. 
We then take the average of these complex-valued points,
$\zavg = \frac{1}{n}\sum_{k=1}^{n} \zeta_k$. 

We next calculate two quantities: 
the average phase difference 
$\Phi = \arg (\zavg) / (2 \pi)$
and the phase order 
$\Omega = |\zavg|$.
The average phase difference $\Phi$ 
is the circular average of the peak-by-peak phase difference
between the two signals through time.
The phase order $\Omega$ tells us 
how concentrated the phase differences are compared to the average. 
If $\zeta_k \approx \zavg$ for all $k$, then $|\zavg| \approx 1$.
However, if the values of $\zeta_k$ are
uniformly spread around the unit circle,
we would have a $\zavg \approx 0$,
since opposite phases cancel out. 
Thus, the phase order $0 \leq \Omega \leq 1$, 
and the closer it is to one,
the more reliable the phase difference is between the two rhythms over time. 

\subsubsection{Irregularity scores}
\label{sec:irregularity}

We define the irregularity score of sequence $x_j$ as
\begin{equation}
  \label{eq:irregularity}
  \mathrm{IRS}(x) = 
  \frac{1}{\nbursts} 
  	\sum_{j=1}^{\nbursts}
    \frac{\left| x_{j+1}-x_j \right|}{\left| x_j \right|} 
    \,.
\end{equation}
Here, $x_j$ denotes either the amplitude of the $j$th detected burst
(amplitude irregularity)
or the period between bursts $j$ and $j+1$ 
(period irregularity).
The irregularity score IRS$\,(x)$ 
measures the average relative change in $x$.

\subsubsection{Statistical tests}

We analyzed the amplitude, period, amplitude irregularity, and period irregularity
using a linear mixed effects model.
This model captures the repeated measurement structure inherent in our 
experimental design.
In particular, we model the response (amplitude, period, etc.)
$y_{s,d}$ of a slice $s$ to drug $d$ as 
\[
	y_{s,d} = a + a_s + \mu_{d} + \epsilon_{s,d},
\]
where $a$ is a fixed intercept (representing the control level of $y$), 
$a_s$ is a zero-mean random effect for each slice,
$\mu_d$ is a fixed effect for each drug (DNQX or DNQX+PTX),
and $\epsilon_{s,d}$ is a zero-mean noise term.
We fit this model using the {\tt lmerTest} package in R, and
the code and data used for fitting and analysis are provided in the Data Supplement.
In the results we report the estimate of the fixed effects ($a$, $\mu_d$),
standard error (SE), degrees of freedom (DF), $t$ value, and p value.

\section{Results}

We developed a network model of the \prebot 
and used this to examine the impact of connectivity and inhibition.  
Each cell in the network is governed by membrane currents that can 
produce square wave bursting via the persistent sodium current $\INaP$ \citep{butera1999}. 
We include bursting pacemaker (B), tonic spiking (TS), 
and quiescent (Q) cell types in realistic proportions.
Through simulations, 
we examine the effects of network connectivity
and the presence of inhibitory cells on rhythm generation.
To achieve this, we vary three key parameters over their biologically plausible ranges:  
(1) the fraction of inhibitory cells $\pI$, 
(2) the average total degree $\kavg$, 
i.e.\ the average total incoming and outgoing connections incident to a neuron, and 
(3) excitatory and inhibitory maximal synaptic conductances $\gE$ and $\gI$.  
The parameter $\kavg$ controls the sparsity of synaptic connections present in the network;
as $\kavg$ increases, the network becomes increasingly connected.

As we detail below, we compute metrics 
of synchronous bursting within the microcircuit as these network parameters vary.
We then generalize the model to two coupled microcircuits
and test whether the added network structure can generate multi-phase rhythms.
Finally, we also compare these model effects to 
experiments with \prebot slice preparations,
where we use a pharmacological approach to
modulate the efficacy of excitatory and inhibitory synapses.

\begin{figure}[t!]
\centering
\includegraphics[width=\linewidth]{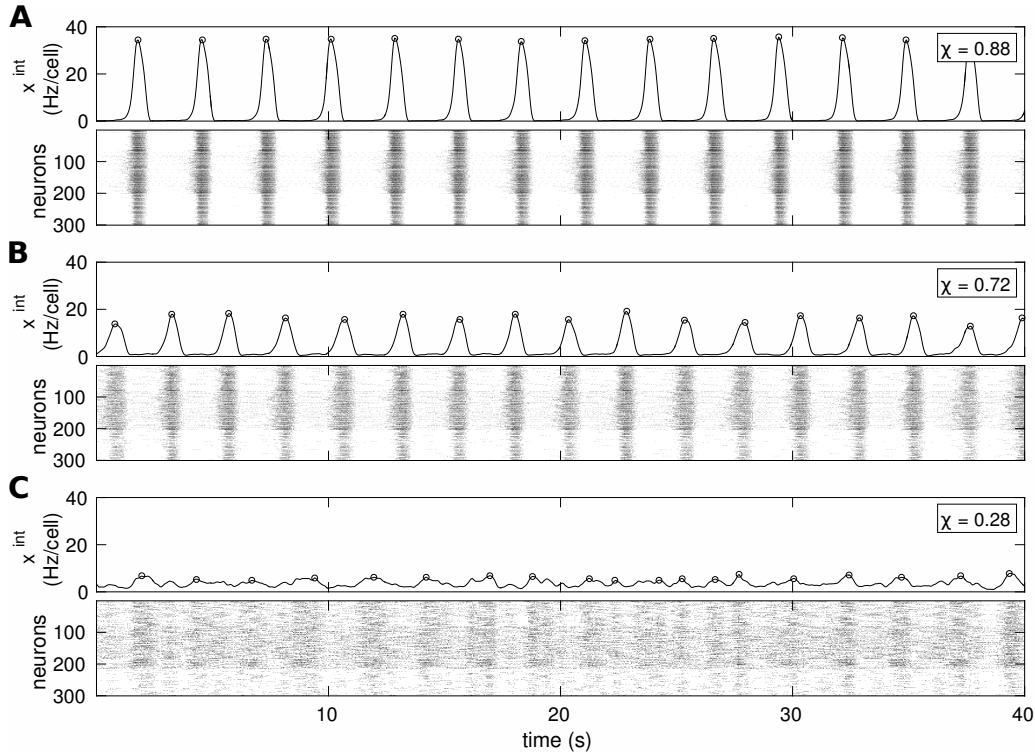}
\caption{
  {\bf With higher fraction of inhibitory cells, 
    synchrony and burst amplitude decrease,
    and the integrated timeseries becomes more variable.}
  Three simulations of the respiratory network model:
  {\bf A}, $\pI=0\%$;
  {\bf B}, $\pI=20\%$;
  {\bf C}, $\pI=40\%$.
  Above, we show the integrated trace, 
  which is a lowpass-filtered average of the spiking activity
  of all $N=300$ neurons in the network.
  Below, we show the spike raster of individual neuron activity.
  In all cases, $\kavg=6$, $\gE=\gI=2.0$ nS. 
  Detected bursts are marked by open circles on the integrated traces.
  At lower levels of synchrony, as in part C,
  what constitutes a burst becomes ambiguous.
}
\label{fig:rasters}
\end{figure}

\subsection{Inhibition and sparsity weaken the model rhythm}
\label{sec:inh_one_pop}

\begin{figure}[tb!]
  \centering
  \includegraphics[width=\textwidth]{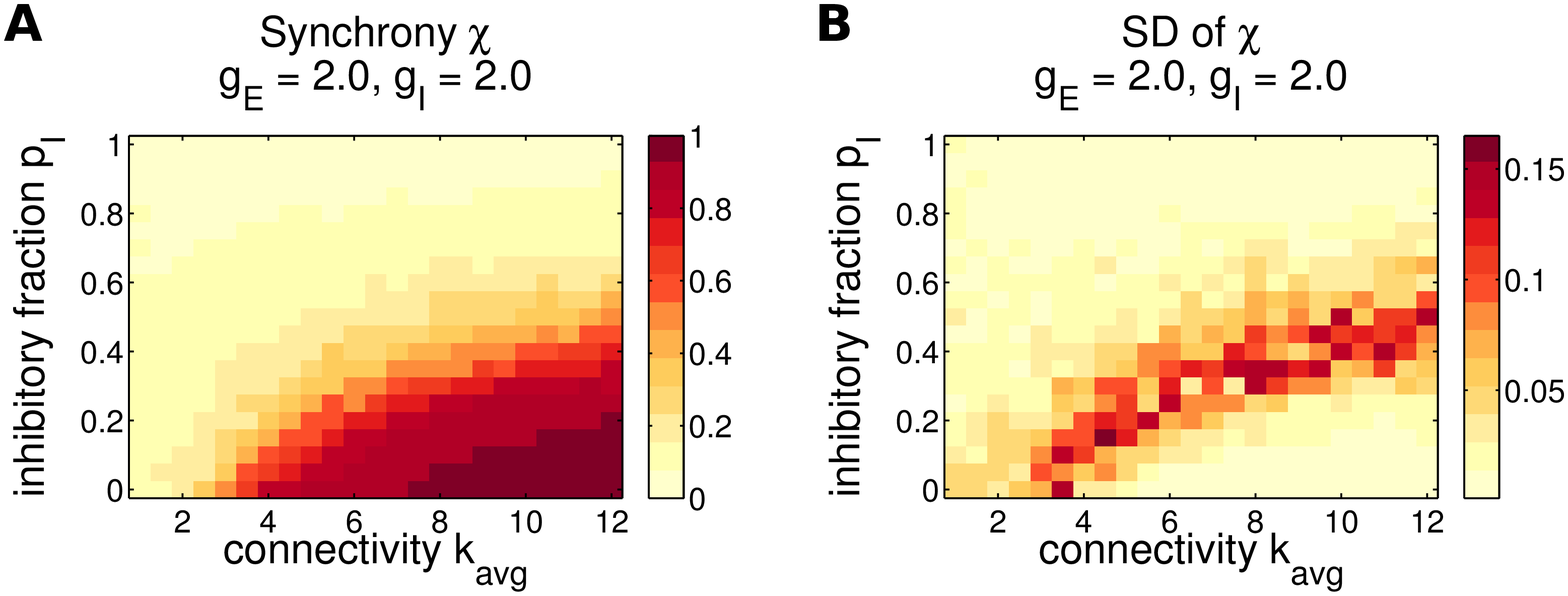}
  \caption{
    {\bf Synchrony decreases with inhibition and sparsity. 
      The highest variability across networks occurs at the 
      synchronization boundary.}
  {\bf A}, Synchrony parameter 
  $\sync$ averaged over 8 network realizations,
  plotted versus the amount of connections $\kavg$
  and the fraction of inhibitory neurons $\pI$.
  {\bf B}, Standard deviation of $\sync$ over
  network realizations.
  Higher standard deviation indicates that the synchrony is not 
  reliable for different networks with those parameters.
  The area of highest standard deviation occurs at the boundary of 
  low and high synchrony, $\sync \approx 0.5$. 
  This is indicative of a phase transition between 
  synchronized and desyncronized states.}
  \label{fig:synchrony}
\end{figure}

We first fix a moderate level of network sparsity, so that each cell receives and sends
a total of $\kavg=6$ connections on average, and we also fix the 
synaptic strengths ($\gE$ and $\gI = 2.0$ nS).
In Fig.~\ref{fig:rasters}, we show the behavior of the
network for varying amounts of 
inhibitory cells $\pI$. 

In Fig.~\ref{fig:rasters}A, 
the inhibitory fraction $\pI=0$,
so the network is purely excitatory.
In this case it generates a strong, regular rhythm, 
and the population is highly synchronized.  
This is clear from both the integrated trace 
$\popint$,
which captures the network average activity and thus the rhythm (defined in Section~\ref{sec:filtering}),
and the individual neuron spikes in a raster, 
which are clearly aligned and periodic across many cells in the microcircuit. 
To further quantify the levels of synchronized firing,
we use the synchrony measure $\sync$, a normalized measure of the 
individual neuron correlations to the population rhythm,
formally defined in Eqn.~\eqref{eq:chi}.
Values of $\sync \approx 1$ reflect a highly-synchronized population,
whereas $\sync \approx 0$ means the population is desynchronized.
The cells in panel A are visibly synchronized from the raster,
and have synchrony $\sync = 0.88$.

We introduce a greater fraction of inhibitory cells $\pI=0.2$ in panel B.  
Here, we see more irregularity in the population rhythm as well as
and reduced burst amplitude and synchrony ($\sync = 0.72$).
In panel C, with a still greater fraction of inhibitory cells, $\pI=0.4$,
the network shows further reduced synchrony ($\sync = 0.28$)
and a very irregular, weak rhythm.
In this case, the ``rhythm'' is extremely weak,
if it even can be said to exist at all, and could not drive healthy breathing.  

Building on these three examples, we next studied the impact of inhibition on synchrony over a wider range of network connectivity parameters.  Here, we vary not only the fraction of inhibitory cells $\pI$, but also the sparsity via $\kavg$.  
In Fig.~\ref{fig:synchrony}, 
we summarize the effects of inhibition and sparsity on synchrony by plotting $\sync$
as those parameters vary.  
Each point in the plot is the average $\sync$ over 8
network realizations with the corresponding parameters.
The main tendency is for higher synchrony with higher $\kavg$,
i.e.\ higher connectivity and less sparsity,
and lower synchrony with higher $\pI$.
A similar effect occurs when varying
$\gE$ and $\gI$,
where stronger excitation synchronizes
and stronger inhibition desychronizes
(shown in Fig.~\ref{fig:irr_score_synaptic} for comparison with pharmacological experiments).

Inhibition thus decreases the synchrony within the \prebot microcircuit,
which hinders the rhythm.
At or above $\pI=50\%$,
the network is desynchronized for all connectivities $\kavg$.
With an inhibitory majority, 
most inputs a neuron receives are desynchronizing,
thus no coherent overall rhythm is possible. 
This is one of our first major results: 
In a single microcircuit, 
constructed with {\it homogeneous} random connectivity
and with $\INaP$-driven burst dynamics \citep{butera1999},
inhibition cannot lead to the creation of a multi-phase rhythm.
Inhibition only has the effect of desynchronizing bursting neurons 
and disabling the overall rhythm.

For any type of random connectivity,
there is no single network corresponding to a given inhibitory fraction and sparsity level.
Rather, each setting of these parameters defines a probability distribution 
over a whole family of networks, and we can study rhythm generation on sample realizations.
This raises the question of how consistent our findings are from one of these networks to the next. 
To address this, we next depict the standard deviation of 
$\sync$
across the 8 network realizations,
shown in Fig.~\ref{fig:synchrony}B.
The standard deviation tells us how much variation in synchrony
to expect for different random networks with these parameters,
with a higher standard deviation indicating less reliability.
The variability in networks is a result of their random generation.
The highest standard deviation occurs
near the border between synchrony and disorder, where
the average
$\sync \approx 0.5$
(see panel A).
Above this border, 
almost all networks exhibit low synchrony,
and below it
networks consistently show the same levels of high synchrony.
Near the transition,
random variations in the network structure
have a larger effect on synchrony.
The increase in standard deviation at the boundary between high and low synchrony
is indicative of a ``phase transition'' between synchronized and 
desynchronized network states \citep{arenas2008}.

\subsection{Inhibition creates an expiratory subpopulation}
\label{sec:anomalous_expir}

\begin{figure}[tb!]
  \centering
  \includegraphics[width=\textwidth]{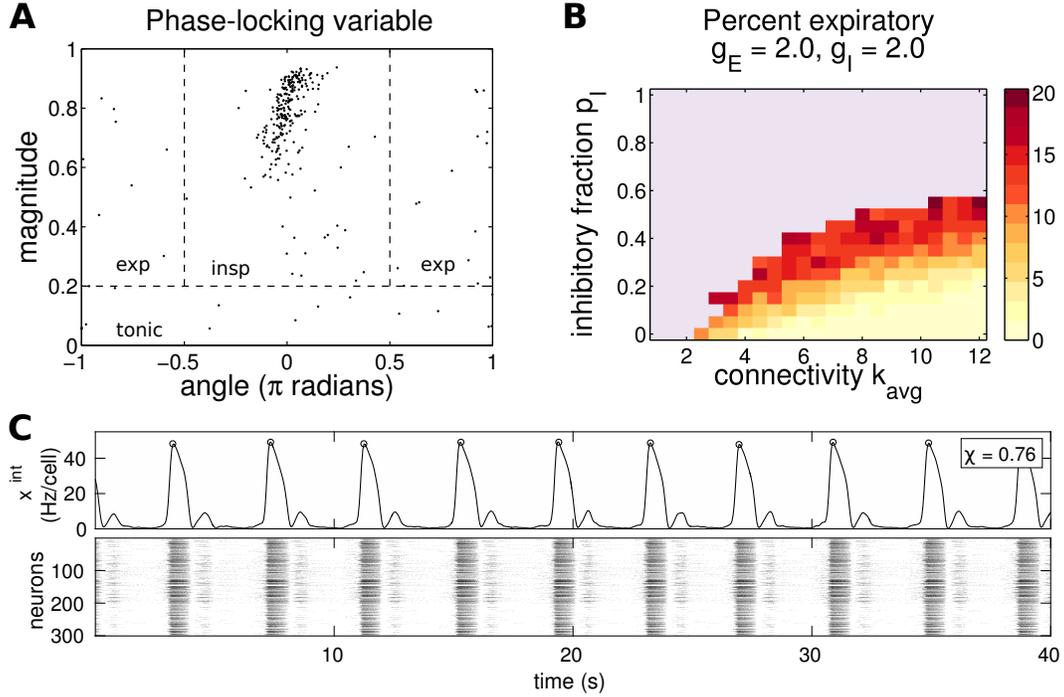}
  \caption{
    {\bf Expiratory cells arise from inhibition, 
      but can only occupy a minority without disrupting the 
      inspiratory rhythm.}
    {\bf A},
    Neuron phase-locking variables 
    for the simulation in Fig.~\ref{fig:rasters}B
    ($\kavg=6$, $\pI=20\%$).
    Each neuron has an associated complex number $z_i$ with $0 \leq |z_i| \leq 1$.
    The magnitude $|z_i|$ is plotted against angle $\arg z_i$.
    These are used to define inspiratory, expiratory,
    and tonic neurons 
    via the labeled regions separated by the dashed lines.
    {\bf B},
    Expiratory (anti-phase with main rhythm) 
    neurons as a function of network parameters 
    $\kavg$ and $\pI$. 
    The fraction of expiratory neurons
    increases with inhibition or as the connectivity becomes weaker.
    The blue indicates the absence of any overall rhythm, 
    defined as $\sync < 0.25$.
    {\bf C}, 
    An example of a simulation with two-phase activity,
    with $\kavg = 6$, $\pI=30\%$, $\gE=5.0$, and $\gI=2.0$.
    A minority of neurons produce a reliable, small bump
    after every burst.
    It is aligned near $0.7 \pi$, so it is more of a 
    post-inspiratory or pre-expiratory burst.
    These expiratory cells are rebound bursting after being 
    disinhibited.
    This is similar to the ``handshake'' mechanism of 
    \citet{wittmeier2008}.
    However, this type of two-phase rhythm is very rare in simulations.
}
  \label{fig:expiratory}
\end{figure}

\begin{figure}[tb!]
  \centering
  \includegraphics[width=\linewidth]{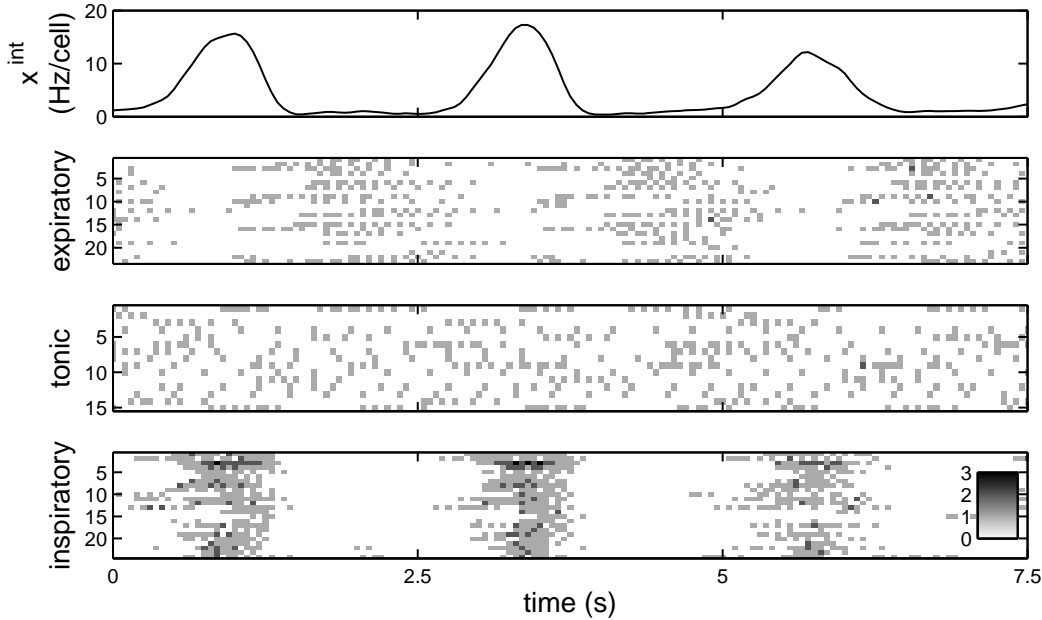}
  \caption{
    {\bf Example rasters of expiratory, tonic, 
    and inspiratory cells.}
    Expiratory cells exhibit lower firing rates than inspiratory ones,
    similar to the typical tonic firing observed in slices.
    As shown, tonic classified cells can be bursting so long as their
    bursts do not 
    occur reliably at any given phase.
    The inspiratory cells shown are a random subset.
    Data are for a representative network with
    $\kavg=6$, $\pI=20\%$ 
    (same as Figs.~\ref{fig:rasters}B
    and \ref{fig:synchrony}B).
  }
  \label{fig:cell_types}
\end{figure}

\begin{figure}[tb!]
  \centering
  \includegraphics[width=\linewidth]{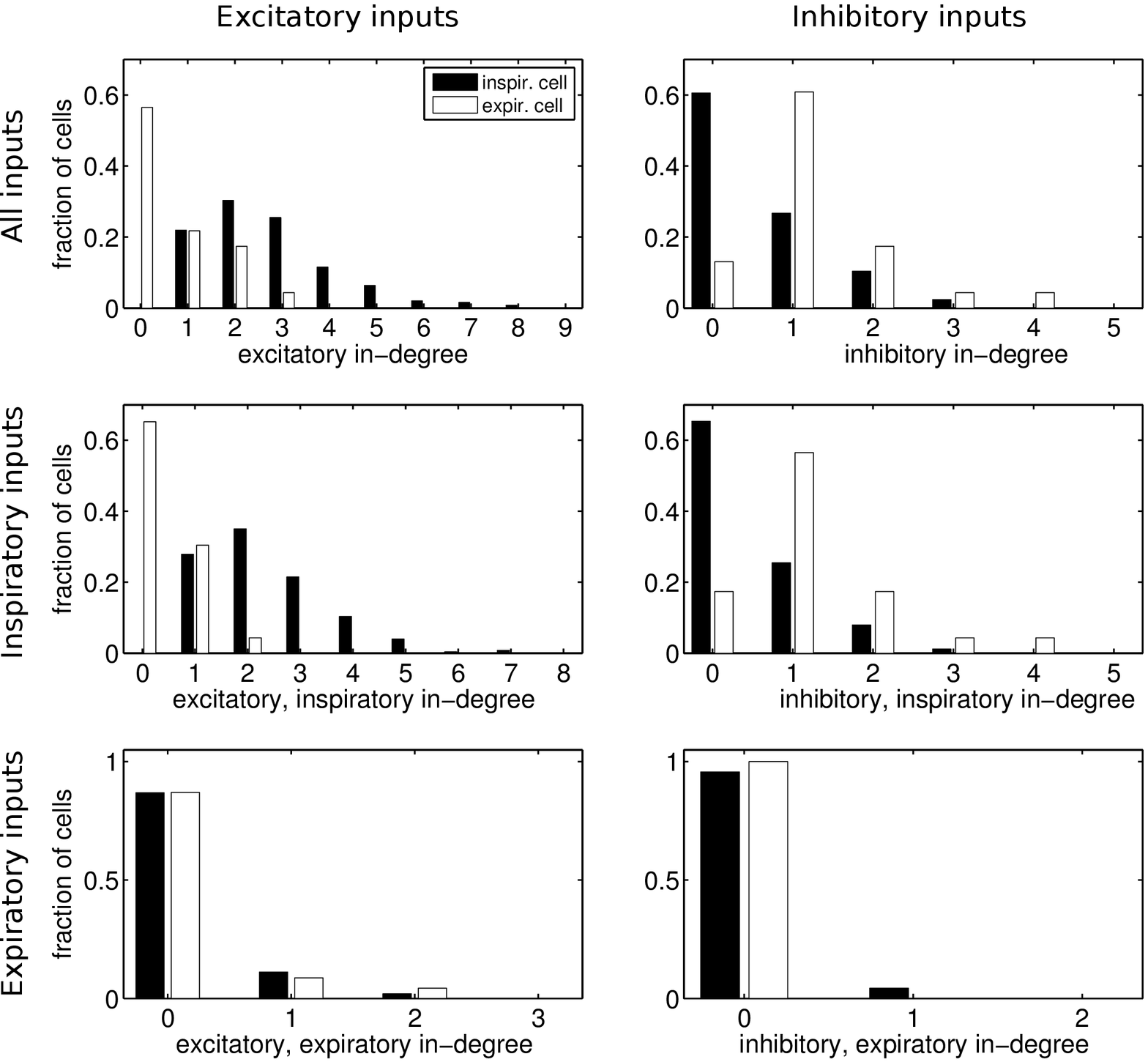}
  \caption{
    {\bf Expiratory cells preferentially receive input from
    other excitatory, expiratory cells 
    and inhibitory, inspiratory cells.}
    {\bf B}, 
    The top row shows the distribution of inputs,
    excitatory on left and inhibitory on right,
    colored by whether the receiving cell is inspiratory (black bars) 
    or expiratory (white bars).
    Expiratory cells receive less excitatory and more inhibitory connections
    than inspiratory cells.
    The center and bottom rows breaks down these inputs
    by the phase of the presynaptic neuron,
    inspiratory inputs shown in the center and expiratory below.
    Expiratory cells preferentially receive excitatory input from
    other expiratory cells (compare middle left and bottom left).
    Furthermore, inhibitory input to expiratory cells
    tends to come from inspiratory cells rather than 
    other expiratory cells (middle right and bottom right).
    Data are for a representative network with
    $\kavg=6$, $\pI=20\%$ 
    (same as Figs.~\ref{fig:rasters}B
    and \ref{fig:synchrony}B).
    There were 251 inspiratory, 23 expiratory, 15 tonic, 
    and 11 silent cells.
  }
  \label{fig:degreehistograms}
\end{figure}

In the \prebot, the majority of cells fire in phase with inspiration, but there are also
cells that fire during other phases (post-inspiratory or expiratory)
along with tonically active cells.
A goal of our study is to identify the network and inhibitory effects leading 
to this variety of cells.

In order to analyze the time during the ongoing population rhythm
at which individual model neurons are active,
we identify robust peaks in the integrated trace as population bursts
(see Section~\ref{sec:phases} for details).
This allows us to map time into a phase variable 
$\phi \in [-\pi, \pi]$
and study neuron activity triggered on phase.
Each peak in the rhythm occurs as the population 
bursts in synchrony and sets the phase $\phi = 0$.
Values of $\phi \approx 0$ correspond to
the inspiratory phase,
since this corresponds to activity in phase with the overall population rhythm,
which for the \prebot is inspiration.
A phase near $\pi$ or $-\pi$ we call expiratory.
We examine cells' firing rates as a function of phase,
which we call the burst triggered average (BTA, Eqn.~\ref{eq:bta}).
Using this, we define a phase-locking variable
$z_i$ (Eqn.~\ref{eq:OP}) for each cell.
The magnitude $|z_i|$ reflects how selectively cell $i$
responds to phase,
and the angle $\arg(z)$ tells the phase it prefers.
This allows us to classify cells as 
inspiratory, expiratory, tonic, or silent.
Fig.~\ref{fig:expiratory}A
shows the phase-locking variables $z_i$ 
for an example simulation with parameters that generate a realistic rhythm
($\kavg=6$, $\pI=20\%$, $\sync=0.716$, 
with raster and integrated trace in
Fig.~\ref{fig:rasters}B).
In this case we see most neurons are inspiratory,
with a dominant cluster of phase-locking variables centered on 
$|z| \approx 0.8$ and $\arg(z) \approx 0$.
The rest of the cells are distributed approximately uniformly
at random in the phase/magnitude cylinder.
In this example, the majority of cells are inspiratory, 
with a smattering of expiratory and tonic cells.

Panel B in Fig.~\ref{fig:expiratory} shows our main results.
For any connectivity level $\kavg$,
we find that the number of expiratory neurons increases
as the fraction of inhibitory cells $\pI$ increases
until the rhythm degrades entirely.
Note that there can be a few expiratory neurons even with
$\pI = 0$
for
$\kavg < 4$.
However, at this connectivity each cell has less than 2 incoming connections on average.
The expiratory cells in that case are isolated 
from the rest of the network and have in-degree zero,
with their phase only reflecting random initial conditions.
Comparing Figs.~\ref{fig:synchrony}A and \ref{fig:expiratory}B,
we see that the number of expiratory neurons grows
as synchrony decreases.

Another key finding of panel \ref{fig:expiratory}B
is that there are never more than 20\% expiratory cells.
This means that, 
in this kind of 
unstructured microcircuit,
it is not possible to create a 
two-phase rhythm
where the expiratory burst is of similar magnitude 
to the inspiratory burst.
Up to approximately $20\%$
of neurons can be expiratory 
without destroying the rhythm,
defined as maintaining $\sync \geq 0.25$.
Fig.~\ref{fig:expiratory}C 
shows an example of a rhythm with two phases,
where the expiratory or post-inspiratory phase 
recruits only a minority of cells.
The expiratory burst in this case is 
caused by rebound bursting of expiratory cells
when they are released from inhibition.
However, a two-phase rhythm of this magnitude is 
rare in our simulations.
For example, it does not occur in other network realizations
with the same parameters as Fig.~\ref{fig:expiratory}C.

One of our goals is to understand the network mechanisms
that give rise to expiratory cells. 
In Fig.~\ref{fig:cell_types}, 
we show the firing properties of some example expiratory,
tonic, and inspiratory classified cells.
Expiratory and tonic cells both fire at lower rates than
inspiratory cells, which are active in tight bursts.
The modeled expiratory cells thus show tonic active
behavior which is suppressed by inhibition,
as observed in slice \citep{shao1997,lieske2000}.
Note that some of the tonic cells in Fig.~\ref{fig:cell_types}
are bursting, just not at a reliable rhythm phase.

Each neuron's phase-locking properties
are determined by its intrinsic dynamics
and the excitatory and inhibitory
synaptic currents it receives during various phases of the rhythm.
In the model, we find that expiratory cells 
receive different synaptic inputs than inspiratory cells.
We can see this by plotting their input properties 
in Fig.~\ref{fig:degreehistograms},
in this case for a typical simulation in the partially synchronized
regime, the same parameters as Fig.~\ref{fig:rasters}B.
Overall, expiratory cells have less excitatory inputs and
more inhibitory inputs than inspiratory cells (top panels).
We also break down these inputs by the phase of the presynaptic cell.
Expiratory cells receive less excitation during the inspiratory phase,
and they similarly receive more inhibition during the inspiratory phase
(center panels).
Given that expiratory cells are the minority, 
the trends for inputs during the expiratory phase are not as strong (bottom panels).
This suggests that expiratory cells emerge from random 
configurations in the network, 
which partitions itself into different phases based on the 
types of interactions in each cell's neighborhood.
Excitatory synapses drive the postsynaptic neuron into 
phase with the presynaptic one, 
while inhibitory synapses drive neurons out of phase.

As we have shown in the preceding two sections, 
the presence of inhibition leads to changes in the population rhythm generated in microcircuits: 
a degradation of the overall population synchrony as well
as an increasing presence of expiratory cells.
The average degree $\kavg$ controls the sparsity of connections in the network,
and lower values also lead to less synchrony.
Moreover, we have shown that cells become expiratory due to the arrival of inhibition
during the the inspiratory phase as well as excitation during the
expiratory phase.    

\subsection{Two population network shows the benefits of half-center inhibition}
\label{sec:inh_two_pop}

\begin{figure}[ht!]
	\centering
	\includegraphics[width=\linewidth]{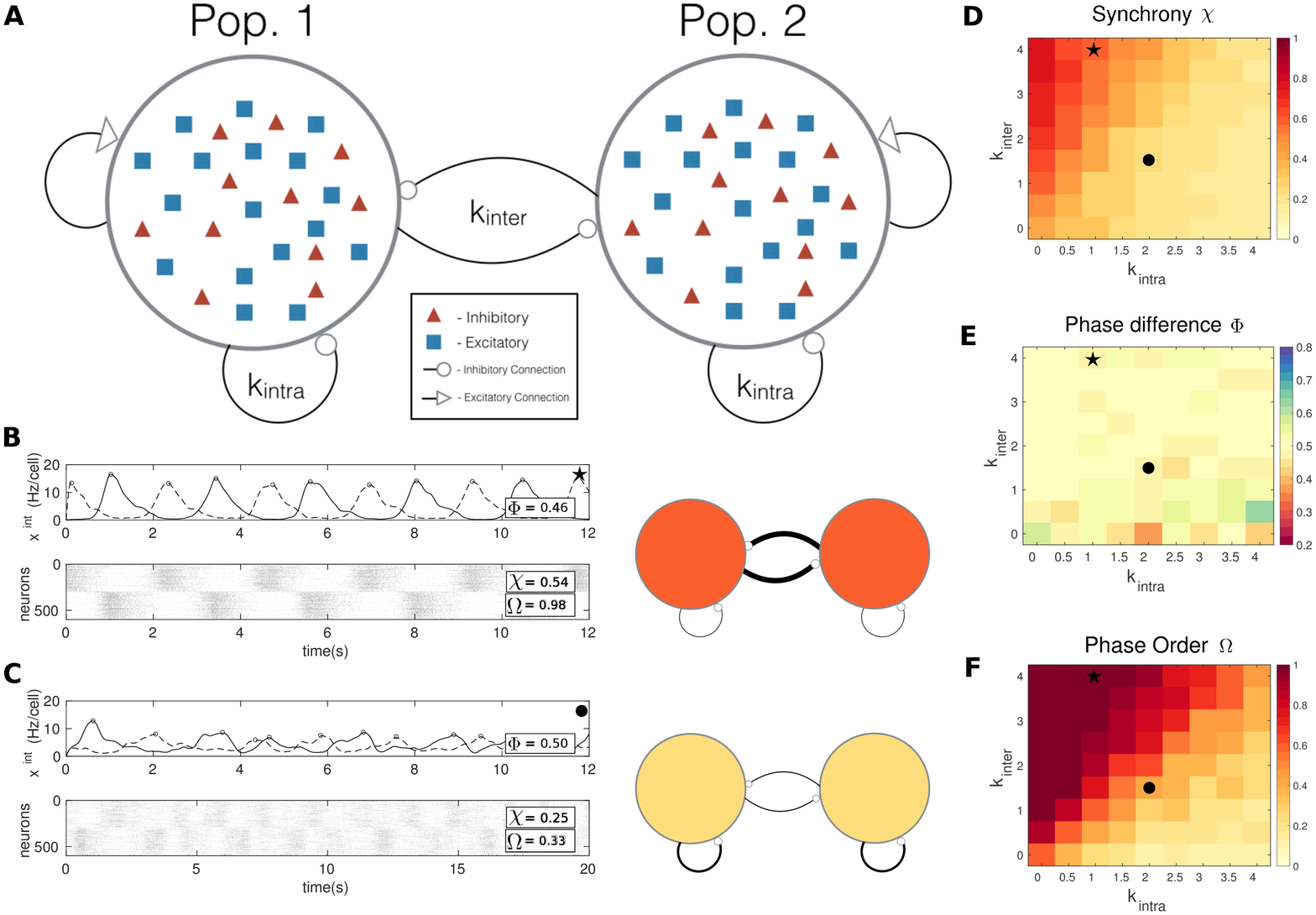}
	\caption{  
    \textbf{A}, Schematic of the two population network.
    The average amount of excitatory connections on average are fixed,
	but we vary the expected
    intra-group and inter-group inhibition $\kintra$ and $\kinter$. 
    \textbf{B} and \textbf{C}, 
    Two simulations of the network with different parameters:
    Each simulation also has a schematic on the right
    demonstrating the differences in inhibitory network strengths. 
    A thicker line indicates more connections, and a 
    darker color indicates a more reliable rhythm.
    Panel B shows the case $\kintra = 1.0$ and $\kinter = 4.0$.
    Panel C depicts $\kintra = 2.0$ and $\kinter = 1.5$. 
   	There, we observe less reliable rhythms, 
   	with decreased phase order $\Omega$ and decreased synchrony $\sync$, 
    despite approximately equal average phase difference $\Phi$. 
    \textbf{D}, 
    Average synchrony over 8 realizations for each $\kinter$ and $\kintra$ pair.
    Higher values of $\sync$ occur above the diagonal $\kinter = \kintra$ line. 
    \textbf{E}, 
    Average phase difference $\Phi$ of rhythmic bursts between the two populations.
    No clear trends are evident, and the value is close to $\Phi = 0.5$, 
    perfectly out-of-phase, in much of the region.
    \textbf{F}, Average phase order $\Omega$. 
    Higher phase order indicates the relative phase of bursts in \popone and \poptwo,
    i.e.\ $\Phi$ in panel E, are reliable.
    The phase order appears to be proportional to the synchrony,
    with the highest values above the diagonal.
    Star and circle symbols in D--F 
    are the network parameters used to produce the rhythms in panels B and C. 
    }
	\label{fig:two_pop_results}
\end{figure}

In Section \ref{sec:inh_one_pop} we examined the effect
of inhibition on rhythmic spiking in a single microcircuit, 
as would model, for example, an isolated \prebot \citep[e.g.][]{ramirez1997}.
There we saw that increasing inhibition causes the synchrony and rhythmicity
of neural spiking to degrade.   
Here, we extend our analysis to a model of two coupled microcircuits. 
Each microcircuit, taken separately, is a heterogeneous subnetwork of cells
with exactly the same properties and parameterization as for the networks studied above.  
The two microcircuits are then coupled with mutual inhibition in the manner of a 
classical half-center pattern generator.  
We explore the effects of inhibition on the synchrony within each microcircuit, 
as well as on the phase of the two microcircuits relative to one another.  

Figure~\ref{fig:two_pop_results}A 
shows a schematic of our network model.  
As in the previous sections,
each microcircuit (a distinct {\it population} of cells) 
contains both excitatory and inhibitory neurons.
For simplicity, since we want to isolate the effects of inhibitory structure,
the excitatory neurons only project locally, that is within the same microcircuit.
We vary inhibitory connectivity via
the parameters $\kinter$ and $\kintra$,
the intra-group and inter-group average degrees 
for inhibitory cells.
For example, setting $\kinter = 0$  yields
independent populations that do not interact; when $\kintra = 0$ and
$\kinter \ne 0$, we have a network version of
the classic half-center oscillator,
with inhibition purely between the two microcircuits.  
We will investigate network activity at these two extremes and intermediate levels of connectivity.  

Panels B and C in Fig.~\ref{fig:two_pop_results} 
illustrate the role of inhibitory connectivity
on rhythmic spiking dynamics in two representative cases.
The upper network (see schematic),  
has weaker inhibition within each population than between the populations,
with parameters $\kintra = 1.0$ and $\kinter = 4.0$.
The population activity exhibits a strong, regular, and
synchronous rhythm with little change in the
phase relationship over time. 
The bottom network has the opposite connectivity:  stronger inhibition within each population
and weaker inhibition between 
($\kintra = 2.0$ and $\kinter = 1.5$).
This network demonstrates a weak, sporadic rhythm with a varying phase
relationship through time. 
These suggest that inhibition within microcircuits competes with inhibition between them to
determine the strength and phase relationships of rhythms. 
We now explore this trend across a broad range of connectivity levels.  

First, we show how intra-\ and inter-group inhibition 
affect the synchrony in the two population model.  
To quantify this, 
we compute the synchrony measures for each population separately 
($\sync_1$ and $\sync_2$), 
and report the average 
$\sync = \left( \sync_1+\sync_2 \right) / 2$.
Figure~\ref{fig:two_pop_results}D
shows the results.
As intra-group inhibition $\kintra$ increases,
there is a degradation in synchrony. 
This is consistent with the results from 
the single population model, 
where unstructured local inhibition reduces the strength and regularity of the population rhythm.  
Panel C gives an example of network activity in this regime,
and is indicated by a circle in panel D--F.
However, as we add inhibitory connections between the two populations by 
increasing $\kinter$,  synchrony recovers: 
overall, we see stronger  synchrony above the diagonal where $\kinter = \kintra$. 
Panel B, indicated by the star in D--F, illustrates this. 
Overall, Figure~\ref{fig:two_pop_results}D suggests that intra-group inhibition destabilizes synchrony, 
while inter-group inhibition can have the opposite effect.

In order to drive breathing, 
in which each microcircuit presumably generates a different phase in a motor pattern,
the model should produce two rhythms with reliable phase separation.
To analyze this, we first compute a measure of the average, 
over time, of the difference between the phases of each microcircuit,
which we call $\Phi$. 
A value $\Phi=1$ or 0 indicates that the two rhythms are,
on average, in-phase, and $\Phi=0.5$ 
indicates the two rhythms are, on average, perfectly out-of-phase
(see further details in methods Sections~\ref{sec:phases} and \ref{sec:phases_2}).
Figure~\ref{fig:two_pop_results}E shows that $\Phi \approx 0.5$ over the range of inhibitory connectivity.  
Thus, the two microcircuits appear to be out of phase on average, regardless of connectivity.
A glance back at panels B and C reveals that this out-of-phase behavior can arise in different ways:  
either for two reliable rhythms that are phase-locked, 
or for two unreliable rhythms that drift broadly with respect to one another over time.
To quantify this difference, we use a phase order metric $\Omega$ (Section~\ref{sec:phases_2}),
shown in Fig.~\ref{fig:two_pop_results}F.
Here, $\Omega = 1$ indicates 
that the phase differences  are completely repeatable over time,
while  $\Omega = 0$ indicates phase differences are completely unreliable, 
instead being evenly spread over time.  
In agreement with the two cases illustrated in panels B and C, 
as we increase the inhibition within microcircuits $\kintra$, 
phase reliability $\Omega$ decreases; 
conversely, increasing $\kinter$ increases $\Omega$. 

These results lead to the important conclusion
that it is not a particular number of inhibitory connections in a network that leads
to a stable two-phase rhythm,
but instead the relative strengths of intra- and inter-group connectivity.
For a stable two-phase rhythm, 
there need to be at least as many inhibitory connections between populations as within populations. 
The key rhythm metrics, 
synchrony $\chi$ and phase order $\Omega$,
demonstrate the same effect,
because $\chi$ and $\Omega$ are strongly correlated.  
This makes sense because the rhythms are generated through synchronous bursting.
Note that an irregularity score for the phase differences
would yield similar results as $\Omega$,
but we prefer $\Omega$ since is takes into the account the circular structure of the phase variable.
Increasing intra-group inhibition pushes the system to the edge of stability.
However, we are able to recover some rhythm stability and phase separation reliability
by increasing inter-group inhibition. 
In summary, we see the same desynchronizing effect of local
inhibition as in the single population model, 
with some benefit to synchronous rhythms possible from inter-group inhibition.

\subsection{Partial synchrony of in vitro \prebot rhythms in multi-array recordings}

\label{sec:partial_sync}

We now turn to experiments with the \prebot,
to test the model predictions about the role of inhibition in such circuits.
We recorded from mouse transverse brainstem slices
containing the \prebot,
keeping only those that initially exhibited robust rhythms.
This yielded a collection of 17 recordings of the population rhythm using a large 
extracellular local field potential (LFP) electrode.
Of these, 4 were simultaneously recorded with a 
linear electrode array to capture the behavior of multiple neurons
(16, 29, 33, and 29 cells were isolated in individual experiments).
From the multi-array data, we extracted individual spikes 
and calculated the synchrony metric $\sync$ as in the model.

\begin{figure}[t!]
  \includegraphics[width=.95\linewidth]{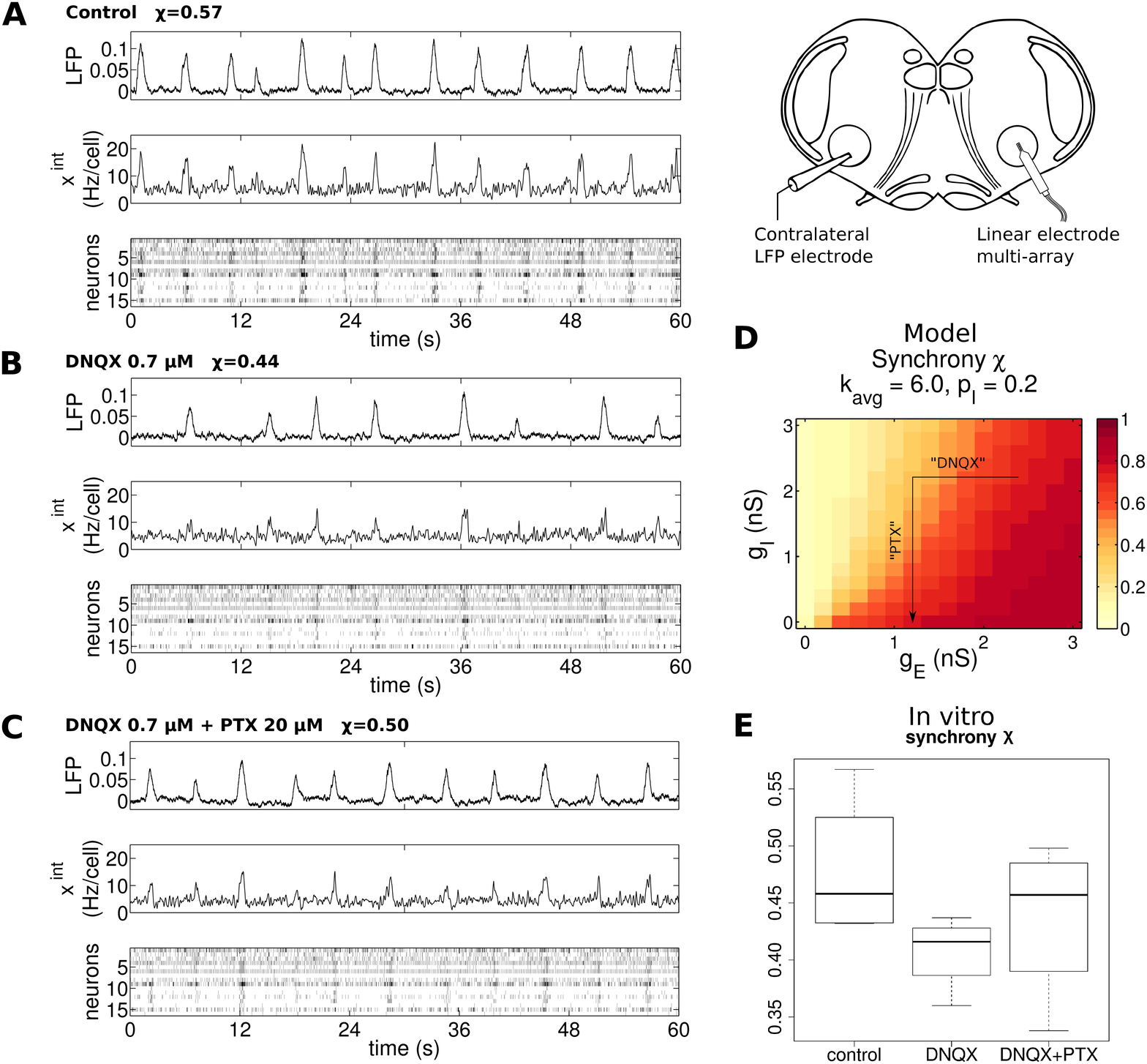}
  \caption{
      {\bf In vitro array recordings from transverse slice preparations
      	exhibit partial synchrony.}
     We performed in vitro \prebot slice experiments, where we measured the
    rhythm in control, partial excitation block (DNQX 0.7~\si{\micro}M), 
    and partial excitation block with full inhibition block 
    (DNQX 0.7~\si{\micro}M + PTX 20~\si{\micro}M).
      We record the \prebot population activity 
      with a large electrode (LFP, arb.\ unit)
      as well as individual neurons in the contralateral area using an array.
      The average activity of the isolated units
      is also shown ($\popint$, Hz/cell).
      {\bf A}, Control conditions show a robust population rhythm
      with some amplitude and period irregularity.
      {\bf B}, Partial excitation block using DNQX degrades the population synchrony,
      with decreased burst amplitude, slower rhythm, and 
      more irregular intervals between bursts.
      {\bf C}, Blocking inhibition with PTX 
      allows the rhythm to recover toward control conditions. 
      {\bf D}, 
      Synchrony in the model, as a function of excitatory and inhibitory
      synaptic conductances $\gE$ and $\gI$,
      increases with stronger excitation and decreases with stronger inhibition,
      similar to varying connectivity $\kavg$ and inhibitory fraction $\pI$.
      Arrows indicate the presumed effects of DNQX and PTX on the model.
      {\bf E}, 
      Measurements of synchrony from our 4 array recording experiments.
      Synchrony takes intermediate values in all conditions, decreasing with DNQX
      and recovering after PTX.
    }   \label{fig:experiment_multi}
\end{figure}

Our experiments reveal that 
a fully synchronized network
such as in Fig.~\ref{fig:rasters}A
is not realistic under our experimental conditions.
This is because
\prebot slices exhibit significant
cycle-cycle variability
\citep{carroll2013,carroll2013a}.
So real networks are somewhere in
the intermediate synchrony range.
We confirmed this in multi-array in vitro experiments.
An example experiment with 16 cells
is shown in Fig.~\ref{fig:experiment_multi}A.
We observe that there is significant cycle-to-cycle 
period and amplitude variability in the rhythm,
which is reflected in the partial synchrony of the
16 neurons recorded ($\sync = 0.57$).
With $n=4$ multi-electrode control experiments, 
we measured an average $\sync = 0.48$ (SD 0.055).

The number of expiratory neurons observed
in other experiments is also consistent
with the degree of partial synchrony in the model.
Multi-array recordings by
\citet{carroll2013}
found
5.0\% expiratory 
and 
3.9\% post-inspiratory 
cells.
Counted together,
as we are doing,
a realistic percentage of expiratory cells is 9\%.
Referring to Figs.~\ref{fig:synchrony}A and \ref{fig:expiratory}B, 
we see that this occurs near the region 
where $\sync \approx 0.6$.
This value is not far from the experimentally 
measured average $\sync = 0.48$.
However, we did not observe any expiratory cells
in our limited set of 4 multi-array experiments,
which is expected based on \citet{carroll2013}.

In Fig.~\ref{fig:experiment_multi}B and C, 
we also show the behavior of the slice under
pharmacological manipulations of the
efficacy of excitatory and inhibitory synaptic transmission,
shown here for completeness and
explored in more detail in Section~\ref{sec:E_I_balance}.
Specifically, we use the glutamatergic antagonist DNQX
and the GABA and glycine receptor antagonist picrotoxin (PTX)
(Section~\ref{sec:experiments}).
After recording the control rhythm,
we applied DNQX 0.7 \si{\micro}M 
to partially block excitation
and observed the resulting rhythm.
After recording in DNQX conditions,
we follow with application of pictrotoxin (PTX) 20 \si{\micro}M.
The dosages are chosen so that DNQX partially blocks 
excitation \citep{honore1988} but does not stop the rhythm,
whereas the PTX dosage is high enough to effect near-complete disinhibition
\citep[see Fig.~1 in][]{othman2012}.
We see in Fig.~\ref{fig:experiment_multi}B that DNQX leads to
less synchrony and a visibly degraded, slower rhythm.
Moreover, Fig.~\ref{fig:experiment_multi}C shows that
when this inhibition is reduced by adding PTX, 
the rhythm recovers toward  
control values of frequency, amplitude, and synchrony.

When varying synaptic conductances in a simulation of the effects of DNQX and PTX,
the computational model behaves as one might expect from our earlier results.
We generated 8 networks with average degree $\kavg = 6$ and inhibitory fraction $\pI = 20 \%$.
Then we varied the maximal conductances of excitatory and inhibitory synapses
$\gE$ and $\gI$ while keeping the network structure fixed.
We show the synchrony $\sync$ as a function of $\gE$ and $\gI$ in 
Fig.~\ref{fig:experiment_multi}D.
Increased $\gE$ leads to enhanced synchrony,
while, as expected from the results above,
increased $\gI$ desynchronizes the population.
Thus, once again we find that excitation synchronizes and 
inhibition desynchronizes activity within a microcircuit.

Finally, in Fig.~\ref{fig:experiment_multi}E we summarize the synchrony $\sync$
across all 4 multi-array experiments and pharmacological conditions.
Clearly, the networks are all partially synchronized. 
Synchrony $\sync$ decreases by about 0.07 (SE 0.02, DF 8, $t$=-3.414, p=0.009) with DNQX, 
with a recovery to near baseline following PTX.
These trends are shown in only 3 out of 4 experiments,
so we stress that this is marginally significant according to the mixed effects model
(see Table~\ref{tab:statistics}).
We next show how proxies for the synchrony which measure regularity
of the rhythm can be applied to our larger collection of LFP recordings to 
further illuminate this trend.

\subsection{Excitatory and inhibitory balance modulates rhythm
irregularity in vitro and in silico}
\label{sec:E_I_balance}

In  Sections \ref{sec:inh_one_pop}--\ref{sec:inh_two_pop}, 
we use a computational model to show how population rhythms 
depend on levels of inhibitory connectivity 
within and between microcircuits.
We have demonstrated that in vitro \prebot networks are naturally
in a partially synchronized state, Sec.~\ref{sec:partial_sync}.
We now investigate how in vitro \prebot rhythms behave under the modulation of
synaptic conductances using pharmacological techniques.
To quantify rhythm quality from the integrated LFP signal,
available in all 17 of our recordings,
we turn to amplitude and period irregularity.
These measure the cycle-to-cycle variability 
of the sequence of burst amplitudes and inter-burst-intervals
\citep[Sec.~\ref{sec:irregularity} and][]{carroll2013,carroll2013a}.

\begin{figure}[t!]
  \centering
	\includegraphics[width=0.9\linewidth]{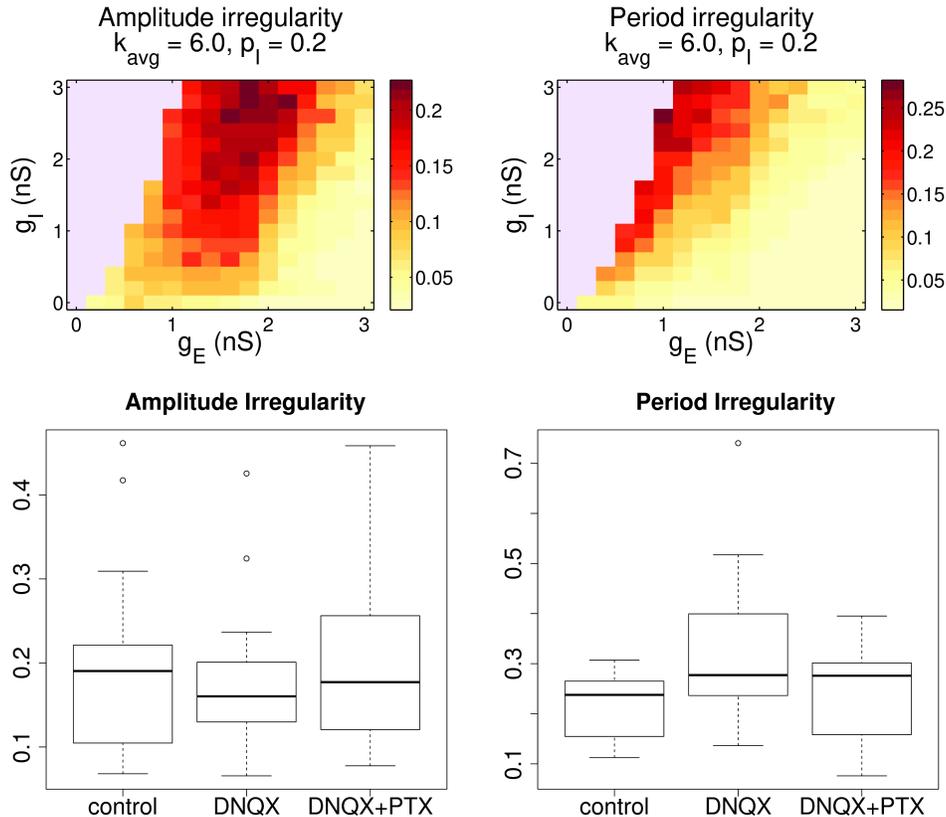}
  \caption{ {\bf Modulation of inhibition and excitation changes the rhythm
      in comparable ways for experiments and the model.}
    {\bf (Above: model)}
    The effect of changing conductances $\gE$ and $\gI$.
    Burst amplitude and period irregularity 
    decrease with stronger excitation and weaker inhibition.
    Both of these measures are negatively correlated
    to the population synchrony, shown in Fig.~\ref{fig:experiment_multi}D.
    {\bf (Below: experiments)}
    This plot summarizes 17 experiments.
    We extracted bursts from the LFP and 
    measured the amplitude and frequency irregularity of those rhythms.
    Amplitude irregularity showed no significant trends across conditions.
    However, period irregularity showed a significant
    increase from control with DNQX,
    a decrease from 
    DNQX to DNQX+PTX,
    and a small increase between
    control and DNQX+PTX.
    See Table~\ref{tab:statistics} 
    for the full output of the statistical tests.
  }
  \label{fig:irr_score_synaptic}
\end{figure}

Our experiments use the synaptic antagonists DNQX and PTX to pharmacologically
modulate the efficacy of excitatory and inhibitory synapses in vitro,
analagous to lowering $\gE$ and $\gI$, respectively.
This is illustrated with the arrows in Fig.~\ref{fig:experiment_multi}D.
In Fig.~\ref{fig:irr_score_synaptic},
we also illustrate the behavior of the amplitude and period irregularity
scores in the model as $\gE$ and $\gI$ vary.
Comparing Figs.~\ref{fig:irr_score_synaptic} and \ref{fig:experiment_multi}D, 
it is apparent that both irregularity scores increase in the model as $\sync$ decreases.
In the 13 experiments where we have only an LFP signal, 
this suggests that irregularity can stand in as a proxy for neuron synchrony,
which we could only measure with multi-cell array recordings.

\begin{table}[tb!]
{ \small
    \begin{center}
	\includegraphics[width=\linewidth]{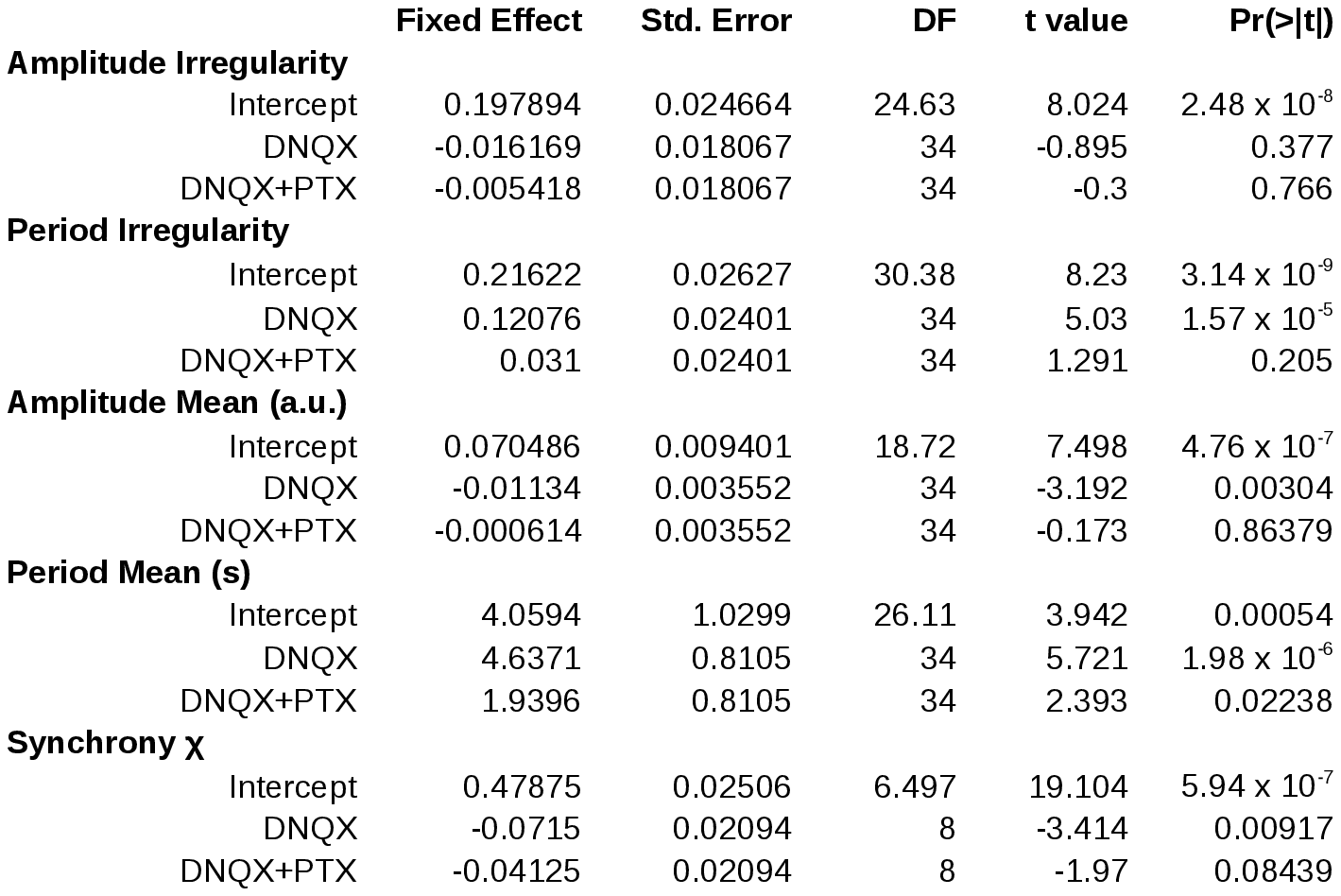}
    \end{center}
    }
    \caption{
    Statistical results for in vitro measurements of
    amplitude irregularity, period irregularity,
    amplitude, and period.
    We report the estimated fixed effect for the intercept, DNQX, and DNQX+PTX conditions,
    as well as standard error (SE), degrees of freedom (DF), $t$ value, and p value for each effect. 
    These data summarize 17 LFP recordings 
    save the synchrony fit, which comes from 4 multielectrode recordings.
     }
    \label{tab:statistics}
\end{table}
      
We plot in vitro irregularity across conditions in 
Fig.~\ref{fig:irr_score_synaptic} using box plots.
The results of statistical tests using a linear mixed effects model
are shown in Table~\ref{tab:statistics}.
To summarize,
amplitude irregularity shows no significant trends with the blocking of excitation via 
DNQX and inhibition via PTX.
However, we noted a statistically significant increase 
(DF=34, t=5.03, p=$1.6 \times 10^{-5}$) in period irregularity
of about 0.12 (SE 0.02)
following application of DNQX and subsequent decrease with PTX
to near baseline. 
The qualitative effect on period irregularity matches trends present in the 
computational network model.

The model also predicts that there would be a
slight decrease in irregularity with initial application of PTX after control, 
i.e.\ a variant of the previous protocol without DNQX.
We performed limited experiments with varying doses of PTX and found some small decreases in
period irregularity which were not significant (data not shown). 
However, it did appear that the more irregular control slices showed greater decreases in irregularity
with application of PTX, 
as also would be expected from the model results in Fig.~\ref{fig:irr_score_synaptic}.

With regards to the lack of a trend in amplitude irregularity, we note that
the ``landscapes'' of the amplitude and period irregularity scores produced by the computational model 
(heat maps in Fig.~\ref{fig:irr_score_synaptic})
show markedly different regions of high irregularity.
In the amplitude irregularity case, the red region of high values is much
wider than in the period irregularity case.
For amplitude, it is shaped like a plateau 
rather than the steep slope of period irregularity.
This suggests that amplitude irregularity is less sensitive to synaptic modulation,
perhaps making trends harder to identify in pharmacological experiments.
However, it could also be that bursting in the real \prebot is 
essentially an ``all-or-nothing'' phenomenon, 
with amplitude irregularity a result of noise but not strongly dependent
on details of the burst dynamics, 
in contrast to the model we study.
This would make it insensitive to blockers, since 
once a burst is triggered it is reliable and consistent, 
similar to the triggering of an action potential.
This is interesting in the context of the burstlet hypothesis \citep{kam2013}.

\subsection{In vitro rhythm slows following excitatory block}
\label{model_problems}

\begin{figure}[t!]
  \centering
  \includegraphics[width=0.9\linewidth]{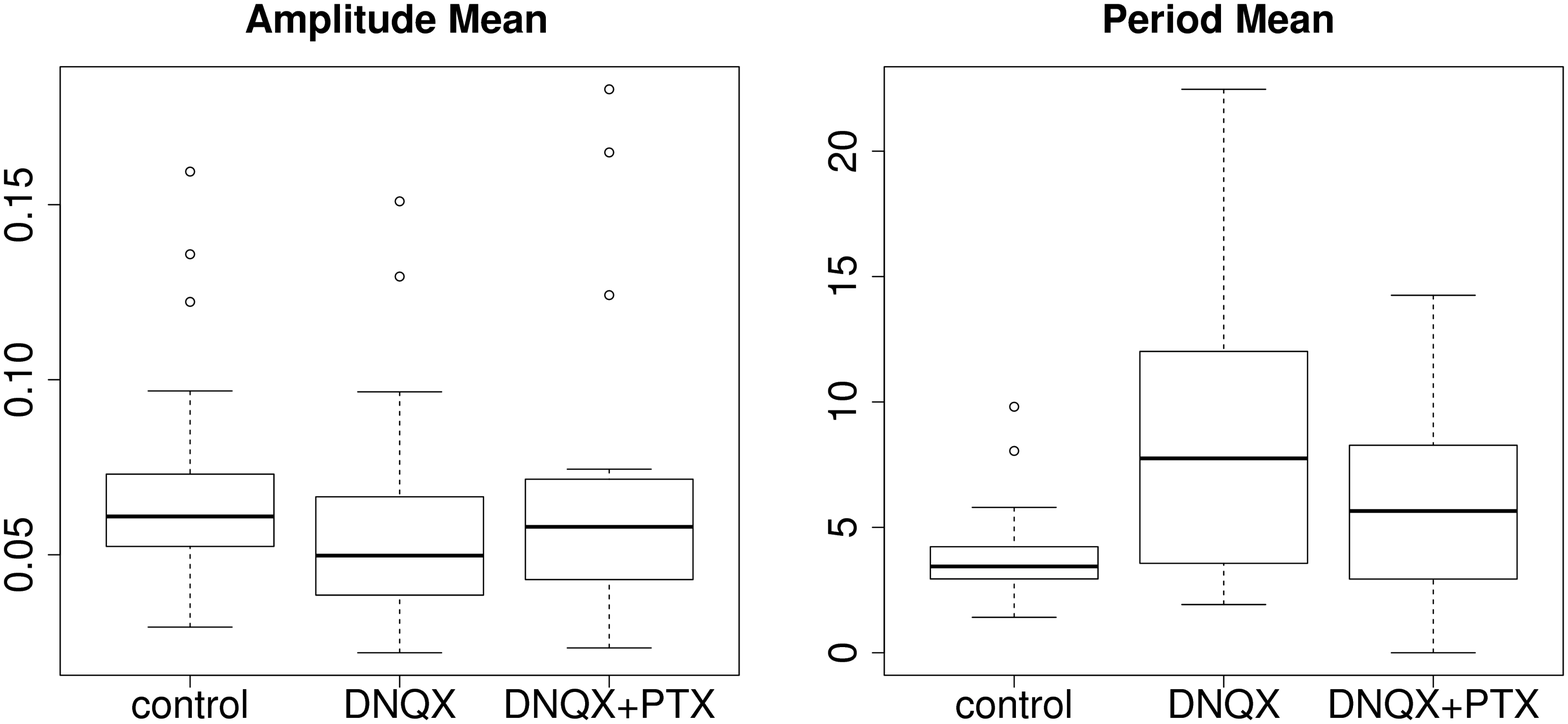}
  \caption{The effect of DNQX and PTX on in vitro
  	rhythm amplitude and period, similar to Figure~\ref{fig:irr_score_synaptic}.
    Amplitude decreases with DNQX while period increases, 
    with both recovering to near baseline after addition of PTX.
    See Table~\ref{tab:statistics} for the result of statistical tests on this data.
  }
  \label{fig:expt_amp_per}
\end{figure}

Besides variability, we found in experiments that 
synaptic blockers also significantly change the overall period
and amplitude of rhythmic bursts, as shown in Fig.~\ref{fig:expt_amp_per} and Table~\ref{tab:statistics}.
Mean burst amplitude
is decreased by -0.011 units (SE 0.004, DF=34, $t$=-3.192, p=0.003)
following DNQX and recovers to baseline with application of PTX.
This is consistent with the effect of varying $\gE$ and $\gI$ in the model.
In experiments, we also see a significant
slowing of the rhythm.
The burst period increases with DNQX by 
4.6 s (SE 0.81, DF=34, $t$=5.72, p=$2 \times 10^{-6}$)
and only partially recovers with application of PTX, 
remaining 1.9 s (SE 0.81, DF=34, $t$=2.39, p=$0.02$) above baseline.
As described above, while our network model qualitatively
predicts the experimental trends for period variability and amplitude modulation in the isolated \prebot,
it does not reproduce overall changes in burst period.

Simple modifications to the model capture the period slowing with excitatory blockers.
Suppose each respiratory cell receives concurrent input from excitatory and inhibitory
pools of tonic neurons \citep{ramirez1997a}.
These cells determine a baseline drive to the \prebot,
which we model as a constant current $\Iapp$.
Tonic external conductances $\gE^{\rm app}$
and $\gI^{\rm app}$ have the same effect but complicate
our parameter tuning due to modification
of the effective leak current.
DNQX would then lower the excitatory drive, 
leading to decreased $\Iapp$.
A negative drive current then slows the amount of time it takes a neuron
to integrate to bursting, 
lowering the neuron's intrinsic burst frequency.
PTX, by lessening the influence of the inhibitory tonic pool, 
causes a net disinhibitory effect on the neuron,
restoring $\Iapp$ to near baseline.
So far, we have taken $\Iapp = 0$ as the baseline,
but these differential effects remain regardless of the
baseline tonic current.
Mimicking DNQX with $\Iapp = -4$ pA causes the period to approximately double
(not shown but tested for $\kavg = 6$, $\pI = 0.2$, $\gE = \gI = 3.0$ in control, 
$\gE = 1.8$ under DNQX).

One consequence of this tonic pool hypothesis is that 
changing the baseline drive also changes the intrinsic dynamics of neurons.
Increased hyperpolarization can cause tonic cells to become bursters, 
and bursters to become silent in the absence of network effects.
However, this can benefit synchrony, 
since when the large pool of originally tonic cells shift into bursting mode,
they can help maintain a strong rhythm despite the reduced excitatory synaptic drive. 
In a check for a few network structures, we found that the `main' effects of excitation and 
inhibition on rhythms persist when we also make these $\Iapp$ changes.

To recap, our experimental results show that the control \prebot
networks lie in the partially-sychronized regime.
The results also confirm that the relative balance of 
excitation and inhibition determine the level of synchrony
and variability of the rhythm.
In experiments, we also find a strong dependence
of rhythm frequency on the amount of inhibition,
and we have discussed changes to the model which could explain this effect.

\section{Discussion}

\subsection{Network structure of respiratory areas}

The \prebot contains neurons which are
silent, tonic spiking, or periodically bursting pacemakers
\citep{thoby-brisson2001,pena2004,ramirez2011}.
Numerous models are proposed for the \prebot,
at the level of single neurons with pacemaker dynamics
\citep{butera1999,best2005,rubin2009a,toporikova2011,park2013}
as well as networks of these neurons
(\citealp{butera1999a, best2005, purvis2006, rubin2009};
\citealp{rubin2009a, schwab2010, gaiteri2011,lal2011, 
rubin2011,carroll2013,carroll2013a,wang2014}).
Traditionally, these models have consisted of just the excitatory, essential
core of inspiratory neurons.
However, \citet{ramirez1997a} showed that inspiratory cells 
receive concurrent excitation and inhibition in the inspiratory phase 
during both in vitro and in vivo recordings from cat \prebot.
Furthermore, \citet{morgado-valle2010}
demonstrated the existence of glycinergic inspiratory pacemakers within \prebot, 
likely candidates for the inhibitory population presynaptic to those found by Ramirez et al.
We have chosen to study the consequences of mixed 
excitatory and inhibitory cells in this network.

The details of network structure in the \prebot is currently unknown,
and molecular markers for rhythmogenic neurons have been found only recently
\citep{wang2014}.
\citet{rekling2000} recorded from pairs of cells and 
estimated that 13\% (3 of 23 pairs) were synaptically connected. 
However, the distance between the connected neurons of the 3 pairs is unknown.
This, along with the small sample size, makes it difficult 
to know whether this connectivity is representative for the entire \prebot.
Moreover, synaptic transmission was not entirely reliable.
Thus, the robustness of these excitatory connections is difficult to assess from 
those exceedingly difficult paired recordings.  
\citet{hartelt2008}
imaged the dendrites and axons of neurons in the area and
found a network with spatially localized, modular structure
similar to a small-world network.
They estimated average neuron degrees were between roughly
2 and 6 \citep{hartelt2008}.
\citet{carroll2013a} recorded from in vitro slice preparations
and argued for roughly 1\% connectivity using cross-correlation
analysis of 10,778 pairs.
The number of cells in the \prebot is estimated to be around
300--600 \citep{wang2014, winter2009, hayes2012, feldman2013},
although this differs significantly with the estimate of 
3000 neurons by \citet{morgado-valle2010}.
This difference is mainly due to varying functional definitions 
of what constitutes a \prebot neuron.
However, our results should not change much with the network size: 
because we parametrize the connectivity by the average degree, 
the in-degree distribution and thus variability of input signal
to a given neuron
(proportional to $\kavg^{-1/2}$) 
will not change significantly.

The exact structure of the \prebot network remains debatable,
but it appears clear that the connectivity is relatively sparse.
Many original models of the isolated \prebot 
assume a fully-connected network, i.e.\ a complete graph 
\citep{butera1999a,purvis2006,rubin2009a}.
\citet{gaiteri2011} studied a variety of different topologies
and their effects on the rhythm.
Random graphs have recently become more popular
\citep{schwab2010,gaiteri2011,lal2011,carroll2013,carroll2013a,wang2014},
however only a few of these studies have looked at 
sparse random networks with average degree
less than 10 \citep{carroll2013,carroll2013a}.
We believe this sparse regime is relevant to the irregularity observed 
in vitro \citep{carroll2013}.

While a clustered connectivity may be present in the \prebot,
where it would have profound effects on rhythm generation \citep{gaiteri2011},
direct evidence for this is limited to the study of \citet{hartelt2008}.
Furthermore, the \prebot is a bilateral rhythm generator with each side
coupled to the other principally by excitatory connections \citep{lieske2006,koizumi2008},
making the two-population model perhaps well-suited for the \prebot. 
There is also evidence for excitatory connections between the 
expiratory and inspiratory centers \citep{onimaru2009,tan2010,huckstepp2015}.
We did try adding a few excitatory projections between the two populations, 
and in our model only a few projections will make the two centers synchronize.
Having predominantly excitatory connections between bilateral \prebot areas
could further stabilize the rhythm.
However, we have chosen to first model the simpler,
sparse but unstructured random connectivity as presented.
We leave a full exploration of such effects to future work.

\subsection{Rhythm patterning by inhibition}

The neural circuits that drive respiration can generate basic rhythms through excitation alone, 
yet they also include strong inhibitory connections both within and between different microcircuits.  
Our aim here is to shine light on the role of this inhibition. 
Through modeling studies that explored thousands of network configurations,
we show that inhibition plays two main roles in excitatory rhythm generators that depend systematically
on the structure of the underlying connectivity.  
Unstructured local inhibition within a single excitatory microcircuit, as for our model of an isolated  \prebot, 
destabilizes rhythmic bursting by preventing the synchronization of excitatory neurons.
This is in contrast to the spiking models where inhibition facilitates synchrony
and relevant, for example, in the gamma oscillation \citep{borgers2003}. 
Within such a single microcircuit with sparse, random, and homogeneous connectivity, 
adding inhibitory cells does not create a robust two-phase rhythm (i.e., inspiration {\it and} expiration). 
However, such inhibition does explain the presence of 
expiratory cells as have been observed experimentally 
\citep{carroll2013,nieto-posadas2014}.
Our pharmacological experiments in the transverse \prebot slice
also support the presence of local inhibition that is destructive to homogeneous synchrony:  
when we first partially block excitation, and then inhibition,
we see that levels of period irregularity first increase and then decrease.

The same qualitative effects of local inhibition persist in a two population 
inspiratory-expiratory model,
suggesting that the synchronizing and desynchronizing roles
of excitation and inhibition within a population persist in more complicated systems.
Moreover, long-range inhibition {\it between} excitatory microcircuits both
stabilizes rhythms locally (reflected in their synchrony) 
and enforces reliable phase separation between microcircuits (phase order),
reminiscent of the concept of the half-center \citep{brown1911,stuart2008,sharp1996}
This suggests twin roles for inhibition:
Within a single microciruit, it reduces synchrony and introduces some out-of-phase cells; 
between populations, it facilitates partitioning of the overall rhythm into different phases. 
As such, inhibition balances against excitation in a way that depends on the on the overall connectivity of the network.

How strongly do the twin roles for inhibition play out in biological circuits for breathing?
Anatomical studies have suggested substantial inhibition within microcircuits, 
and recordings have shown some cells with expiratory or post-inspiratory 
firing within the predominantly inspiratory \prebot 
\citep{carroll2013,morgado-valle2010,nieto-posadas2014}.  
Intriguingly, our model predicts that the level of local inhibition that is consistent 
with these observations moves the circuits as a whole toward the boundary between 
ordered, synchronous and disordered, asynchronous activity.  
This could be useful for making the network more sensitive to control signals.
For instance, descending excitatory inputs that selectively target the inhibitory population
could lead to pauses in the rhythm.

This frames two questions: First, what constructive role could such destabilizing inhibition play?
Possibly, it could produce a rhythm that has a particular temporal pattern (e.g.\ ramping)
or that could be more flexibly controlled. 
Second, what role might destabilizing inhibition play in disease states in which rhythms 
within and between respiratory and other centers degrade?

Physiological studies suggest interesting answers to the first of these questions.  
Local inhibition within the \prebot has a critical role in shaping the inspiratory pattern
\citep{janczewski2013,sherman2015}, as our modeling study also shows. 
One of the hallmarks of ``eupnea'' or normal breathing is an augmenting ramp-like inspiration
which is lost when inhibition is blocked in the isolated \prebot \citep{lieske2000}.
Characterizing the synaptic profile of inspiratory neurons reveals the presence of concurrent inhibition
and excitation which likely prevent an effective synchronizing between the excitatory neurons, thereby 
slowing down the build-up of inspiratory activity. 
Indeed, we hypothesize that the presence of local, desynchronizing role of inhibition within the \prebot 
could also explain an ongoing debate in the field of respiration, 
i.e.\ why an isolated \prebot can generate a eupnea-like
inspiratory activity pattern in the absence of the other phases of respiratory activity 
\citep{lieske2000,ramirez2003}.
The augmenting inspiratory discharge in the isolated \prebot is very sensitive to the blockade of inhibition. 
In hypoxia, when synaptic inhibition is suppressed, the desynchronizing effect of local inhibition is
lost and the isolated \prebot generates an inspiratory burst that is characterized by a
fast rise time reflective of a facilitated synchronization.
However, the Butera model we implemented does not exhibit these rise time effects at the single-cell level.
Instead, the behavior only becomes evident in the population due to the misalignment of individual
neuron bursts, and this overall effect is quite weak (data not shown).
It is likely that other currents
are important for the individual burst characteristics
and that future models including these will provide further evidence
for a role for local inhibition in shaping inspiratory bursts.

\subsection{Limitations of our study}

There was considerable variability in 
the control rhythms and the responses to drugs.
We believe this is principally due to
intrinsic variability of the \prebot network structure across mice, 
the slicing procedure which damages the network to varying degrees, 
and the moderate dose of DNQX.
The multi-electrode recordings captured between 16 and 33 units.
This small sample of cells contributes significant variance to our synchrony
measure $\sync$, and we believe this is why we cannot see a significant effect on synchrony.
We placed the electrode array where we could record from many inspiratory cells,
however we also found almost as many tonic cells.
It is possible that these are cells which are not integrated into the network
and therefore could bias $\sync$ to lower values.
In future work, it would be important to see whether the rhythm also degrades 
with inhibitory agonists, e.g.\ muscimol \citep[see][]{janczewski2013}.
However, agonists introduce a tonic input
which is rather different than modifying the synaptic efficacies, 
thus they will have a different effect than antagonists or optogenetic stimulation.

Our slice experiments showed a slowing down with excitation block 
and no statistically significant variation in amplitude irregularity,
both in contrast to the model.
Other membrane currents may explain these salient features of our pharmacological studies.
We proposed that tonic populations could drive the change in frequency.
However, the CAN current 
is another likely candidate.
Since CAN-dependent pacemakers can rely on
accumulation of excitatory synaptic events to initiate bursting
\citep{rubin2009,delnegro2010},
excitatory synaptic block will slow this accumulation,
leading to an increase the rhythm period.
This mechanism would be similar to the synaptic integrator
model of \citet{guerrier2015},
which reproduced the period effects of NBQX (similar to DNQX).
As mentioned above, the CAN current is also probably important for 
generation of augmenting, ramping discharges.
Our model excluded CAN for simplicity and because the vast majority of 
respiratory models use the \citet{butera1999}
persistent sodium equations.
Also, it appears that cadmium-sensitive intrinsic bursting neurons
(presumably the same as CAN-dependent) are only a minority
of the respiratory neurons in the \prebot \citep{pena2004}.
\citet{hayes2008} present evidence that a low-threshold, inactivating K$^+$ current
$I_\mathrm{A}$ is present in \prebot neurons and significantly affects rhythmogenesis.
They conclude that $I_\mathrm{A}$ helps control amplitude and frequency irregularity
by preventing or delaying those neurons from responding without massive excitatory input.
Beyond irregularity, $I_\mathrm{A}$ and $I_\mathrm{CAN}$ are
also important for overall burst shape, duration,
inter-spike intervals, burstiness, etc.,
which are interesting topics for future study.
Finally, synaptic delays can be very important determinants of synchronization strength
and phase relationships \citep{brunel1999}.
Future models will need to investigate how these many currents interact with
excitatory and inhibitory synaptic dynamics in rhythm generation.

\subsection{Conclusions}

Our results contribute to a large body of modeling and experimental work in the field.
Because local inhibition has a desynchronizing role, 
the \prebot cannot generate a two-phase rhythm,
consistent with lesioning experiments performed by \citet{smith2007}.
Multiarray recordings from more than 900 neurons that indicate less than 9\% of the neurons in the
\prebot are expiratory \citep{carroll2013} also support this finding. 
Moreover, our modeling study also provides theoretical support for the respiratory network organization
recently proposed by \citet{anderson2016}.
They propose that each phase of the respiratory rhythm is generated by its own excitatory microcircuit 
located in a different region of the ventral respiratory group,
the inspiratory phase being generated by the \prebot, post-inspiration by its own complex 
(the \pico) \citep{anderson2016},
and active expiration by the so-called lateral parafacial/retrotrapezoidal group 
\citep{janczewski2006, onimaru2009, huckstepp2016}.
This idea is similar in spirit to the microcircuit models of
\citet{smith2013,molkov2013,koizumi2013,onimaru2015}, which contain more areas.
However, each of these excitatory microcircuits contains neurons with different anatomical, 
physiological and modulatory properties,
and each is dependent on excitatory synaptic transmission, 
able to generate rhythmicity in the absence of synaptic inhibition
\citep{ramirez2016}.

Overall, a modular organization of rhythm generating networks has both evolutionary \citep{ramirez2016} and functional implications;
the latter may explain, for example, why we can hop on one leg without requiring a major network reconfiguration. 
We hypothesize that the separation of a rhythmic behavior
into several excitatory microcircuits may indeed be dictated
by the architecture of these sparsely connected excitatory networks 
that generate rhythmicity based on excitatory synaptic mechanisms. 
The addition of local inhibition to each microcircuit adds another layer of complexity to 
the generation of rhythms which can affect synchrony and controllability.
The lessons learned from the respiratory circuit may also apply to networks that generate
locomotion or other rhythmic behaviors,
where each phase may be composed of separate microcircuits that are interacting with inhibitory connections.

\section{Acknowledgements}

We would like to thank Jonathan Rubin, Peter J.\ Thomas, 
Juan Restrepo, Bard Ermentrout, and the reviewers
for discussions and suggestions.
KD Harris was supported by a Boeing fellowship and an
NIH Big Data for Genomics and Neurosciences Training Grant.  
J Mendoza was supported by a postbaccalaureate fellowship from the 
University of Washington Institute for Neuroengineering.
J-M Ramirez, FAJ Garcia III, J Mendoza, and T Dashevskiy
were supported by NIH grants 
R01 HL126523 and P01 HL090554.
We also acknowledge support of a NSF Career Award DMS-1056125 to E Shea-Brown
and NIH F32 HL121939 to T Dashevskiy.
This work was facilitated though the use of advanced computational, 
storage, and networking infrastructure provided by the 
Hyak supercomputer system at the University of Washington.

\bibliographystyle{jneurosci}
\bibliography{library}

\begin{thebibliography}{}

\bibitem[\protect\citeauthoryear{Ainsworth  \bgroup et al.\egroup
  }{2012}]{ainsworth2012}
Ainsworth M, Lee S, Cunningham MO, Traub RD, Kopell NJ, Whittington MA (2012)
\newblock Rates and rhythms: A synergistic view of frequency and temporal
  coding in neuronal networks.
\newblock {\em Neuron}~75:\mbox{572--583}.

\bibitem[\protect\citeauthoryear{Anderson  \bgroup et al.\egroup
  }{2016}]{anderson2016}
Anderson TM, Garcia AJ, Baertsch NA, Pollak J, Bloom JC, Wei AD, Rai KG,
  Ramirez JM (2016)
\newblock A novel excitatory network for the control of breathing.
\newblock {\em Nature}~536:\mbox{76--80}.

\bibitem[\protect\citeauthoryear{Arenas  \bgroup et al.\egroup
  }{2008}]{arenas2008}
Arenas A, D{\'\i}az-Guilera A, Kurths J, Moreno Y, Zhou C (2008)
\newblock Synchronization in complex networks.
\newblock {\em Physics Reports}~469:\mbox{93--153}.

\bibitem[\protect\citeauthoryear{Best  \bgroup et al.\egroup }{2005}]{best2005}
Best J, Borisyuk A, Rubin J, Terman D, Wechselberger M (2005)
\newblock The {{Dynamic Range}} of {{Bursting}} in a {{Model Respiratory
  Pacemaker Network}}.
\newblock {\em SIAM Journal on Applied Dynamical Systems}~4:\mbox{1107--1139}.

\bibitem[\protect\citeauthoryear{Bollob{\'a}s}{1998}]{bollobas1998}
Bollob{\'a}s B (1998)
\newblock Random {{Graphs}}
\newblock In {\em Modern {{Graph Theory}}}, number 184 in Graduate Texts in
  Mathematics, \mbox{pp. 215--252}. {Springer New York}.

\bibitem[\protect\citeauthoryear{B{\"o}rgers and Kopell}{2003}]{borgers2003}
B{\"o}rgers C, Kopell N (2003)
\newblock Synchronization in networks of excitatory and inhibitory neurons with
  sparse, random connectivity.
\newblock {\em Neural computation}~15:\mbox{509--538}.

\bibitem[\protect\citeauthoryear{Brown}{1911}]{brown1911}
Brown TG (1911)
\newblock The {{Intrinsic Factors}} in the {{Act}} of {{Progression}} in the
  {{Mammal}}.
\newblock {\em Proceedings of the Royal Society of London. Series B, Containing
  Papers of a Biological Character}~84:\mbox{308--319}.

\bibitem[\protect\citeauthoryear{Brunel and Hakim}{1999}]{brunel1999}
Brunel N, Hakim V (1999)
\newblock Fast {{Global Oscillations}} in {{Networks}} of
  {{Integrate}}-and-{{Fire Neurons}} with {{Low Firing Rates}}.
\newblock {\em Neural Computation}~11:\mbox{1621--1671}.

\bibitem[\protect\citeauthoryear{Butera \bgroup et al.\egroup
  }{1999a}]{butera1999}
Butera RJ, Rinzel J, Smith JC (1999a)
\newblock Models of respiratory rhythm generation in the pre-{{B{\"o}tzinger}}
  complex. {{I}}. {{Bursting}} pacemaker neurons.
\newblock {\em Journal of neurophysiology}~82:\mbox{382--397}.

\bibitem[\protect\citeauthoryear{Butera \bgroup et al.\egroup
  }{1999b}]{butera1999a}
Butera RJ, Rinzel J, Smith JC (1999b)
\newblock Models of respiratory rhythm generation in the pre-{{B{\"o}tzinger}}
  complex. {{II}}. {{Populations}} of coupled pacemaker neurons.
\newblock {\em Journal of Neurophysiology}~82:\mbox{398--415}.

\bibitem[\protect\citeauthoryear{Buzsaki}{2006}]{buzsaki2006}
Buzsaki G (2006)
\newblock {\em Rhythms of the {{Brain}}}
\newblock {Oxford University Press}.

\bibitem[\protect\citeauthoryear{Carroll and Ramirez}{2013}]{carroll2013a}
Carroll MS, Ramirez JM (2013)
\newblock Cycle-by-cycle assembly of respiratory network activity is dynamic
  and stochastic.
\newblock {\em Journal of Neurophysiology}~109:\mbox{296--305}.

\bibitem[\protect\citeauthoryear{Carroll \bgroup et al.\egroup
  }{2013}]{carroll2013}
Carroll MS, Viemari JC, Ramirez JM (2013)
\newblock Patterns of inspiratory phase-dependent activity in the in vitro
  respiratory network.
\newblock {\em Journal of Neurophysiology}~109:\mbox{285--295}.

\bibitem[\protect\citeauthoryear{Cui  \bgroup et al.\egroup }{2016}]{cui2016}
Cui Y, Kam K, Sherman D, Janczewski WA, Zheng Y, Feldman JL (2016)
\newblock Defining {{preB{\"o}tzinger Complex Rhythm}}- and
  {{Pattern}}-{{Generating Neural Microcircuits In~Vivo}}.
\newblock {\em Neuron}~91:\mbox{602--614}.

\bibitem[\protect\citeauthoryear{Del~Negro  \bgroup et al.\egroup
  }{2010}]{delnegro2010}
Del~Negro CA, Hayes JA, Pace RW, Brush BR, Teruyama R, Feldman JL (2010)
\newblock Synaptically {{Activated Burst}}-{{Generating Conductances Underlie}}
  a {{Group}}-{{Pacemaker Mechanism}} for {{Respiratory Rhythm Generation}} in
  {{Mammals}}.
\newblock {\em Progress in Brain Research}~187:\mbox{111--136}.

\bibitem[\protect\citeauthoryear{Del~Negro  \bgroup et al.\egroup
  }{2005}]{delnegro2005}
Del~Negro CA, Morgado-Valle C, Hayes JA, Mackay DD, Pace RW, Crowder EA,
  Feldman JL (2005)
\newblock Sodium and {{Calcium Current}}-{{Mediated Pacemaker Neurons}} and
  {{Respiratory Rhythm Generation}}.
\newblock {\em The Journal of Neuroscience}~25:\mbox{446--453}.

\bibitem[\protect\citeauthoryear{Destexhe \bgroup et al.\egroup
  }{1994}]{destexhe1994}
Destexhe A, Mainen ZF, Sejnowski TJ (1994)
\newblock Synthesis of models for excitable membranes, synaptic transmission
  and neuromodulation using a common kinetic formalism.
\newblock {\em Journal of Computational Neuroscience}~1:\mbox{195--230}.

\bibitem[\protect\citeauthoryear{Feldman \bgroup et al.\egroup
  }{2013}]{feldman2013}
Feldman JL, Del~Negro CA, Gray PA (2013)
\newblock Understanding the {{Rhythm}} of {{Breathing}}: {{So Near}}, {{Yet So
  Far}}.
\newblock {\em Annual Review of Physiology}~75:\mbox{423--452}.

\bibitem[\protect\citeauthoryear{Ferguson  \bgroup et al.\egroup
  }{2015}]{ferguson2015}
Ferguson KA, Njap F, Nicola W, Skinner FK, Campbell SA (2015)
\newblock Examining the limits of cellular adaptation bursting mechanisms in
  biologically-based excitatory networks of the hippocampus.
\newblock {\em Journal of Computational Neuroscience}~39:\mbox{289--309}.

\bibitem[\protect\citeauthoryear{Gaiteri and Rubin}{2011}]{gaiteri2011}
Gaiteri C, Rubin JE (2011)
\newblock The {{Interaction}} of {{Intrinsic Dynamics}} and {{Network
  Topology}} in {{Determining Network Burst Synchrony}}.
\newblock {\em Frontiers in Computational Neuroscience}~5.

\bibitem[\protect\citeauthoryear{Golomb}{2007}]{golomb2007}
Golomb D (2007)
\newblock Neuronal synchrony measures.
\newblock {\em Scholarpedia}~2:\mbox{1347}.

\bibitem[\protect\citeauthoryear{Gray  \bgroup et al.\egroup }{2001}]{gray2001}
Gray PA, Janczewski WA, Mellen N, McCrimmon DR, Feldman JL (2001)
\newblock Normal breathing requires {{preB{\"o}tzinger}} complex neurokinin-1
  receptor-expressing neurons.
\newblock {\em Nature Neuroscience}~4:\mbox{927--930}.

\bibitem[\protect\citeauthoryear{Grillner}{2006}]{grillner2006}
Grillner S (2006)
\newblock Biological {{Pattern Generation}}: {{The Cellular}} and
  {{Computational Logic}} of {{Networks}} in {{Motion}}.
\newblock {\em Neuron}~52:\mbox{751--766}.

\bibitem[\protect\citeauthoryear{Grillner and Jessell}{2009}]{grillner2009}
Grillner S, Jessell TM (2009)
\newblock Measured motion: Searching for simplicity in spinal locomotor
  networks.
\newblock {\em Current Opinion in Neurobiology}~19:\mbox{572--586}.

\bibitem[\protect\citeauthoryear{Guerrier  \bgroup et al.\egroup
  }{2015}]{guerrier2015}
Guerrier C, Hayes JA, Fortin G, Holcman D (2015)
\newblock Robust network oscillations during mammalian respiratory rhythm
  generation driven by synaptic dynamics.
\newblock {\em Proceedings of the National Academy of Sciences}~\mbox{p.
  201421997}.

\bibitem[\protect\citeauthoryear{Hartelt  \bgroup et al.\egroup
  }{2008}]{hartelt2008}
Hartelt N, Skorova E, Manzke T, Suhr M, Mironova L, K{\"u}gler S, Mironov S
  (2008)
\newblock Imaging of respiratory network topology in living brainstem slices.
\newblock {\em Molecular and Cellular Neuroscience}~37:\mbox{425--431}.

\bibitem[\protect\citeauthoryear{Hayes  \bgroup et al.\egroup
  }{2008}]{hayes2008}
Hayes JA, Mendenhall JL, Brush BR, Del~Negro CA (2008)
\newblock 4-{{Aminopyridine}}-sensitive outward currents in
  {{preB{\"o}tzinger}} complex neurons influence respiratory rhythm generation
  in neonatal mice.
\newblock {\em The Journal of Physiology}~586:\mbox{1921--1936}.

\bibitem[\protect\citeauthoryear{Hayes \bgroup et al.\egroup
  }{2012}]{hayes2012}
Hayes JA, Wang X, Negro CAD (2012)
\newblock Cumulative lesioning of respiratory interneurons disrupts and
  precludes motor rhythms in vitro.
\newblock {\em Proceedings of the National Academy of
  Sciences}~109:\mbox{8286--8291}.

\bibitem[\protect\citeauthoryear{Holland \bgroup et al.\egroup
  }{1983}]{holland1983}
Holland PW, Laskey KB, Leinhardt S (1983)
\newblock Stochastic blockmodels: {{First}} steps.
\newblock {\em Social Networks}~5:\mbox{109--137}.

\bibitem[\protect\citeauthoryear{Honore  \bgroup et al.\egroup
  }{1988}]{honore1988}
Honore T, Davies SN, Drejer J, Fletcher EJ, Jacobsen P, Lodge D, Nielsen FE
  (1988)
\newblock Quinoxalinediones: Potent competitive non-{{NMDA}} glutamate receptor
  antagonists.
\newblock {\em Science}~241:\mbox{701--703}.

\bibitem[\protect\citeauthoryear{Huckstepp  \bgroup et al.\egroup
  }{2015}]{huckstepp2015}
Huckstepp RTR, Cardoza KP, Henderson LE, Feldman JL (2015)
\newblock Role of {{Parafacial Nuclei}} in {{Control}} of {{Breathing}} in
  {{Adult Rats}}.
\newblock {\em Journal of Neuroscience}~35:\mbox{1052--1067}.

\bibitem[\protect\citeauthoryear{Huckstepp  \bgroup et al.\egroup
  }{2016}]{huckstepp2016}
Huckstepp RT, Henderson LE, Cardoza KP, Feldman JL (2016)
\newblock Interactions between respiratory oscillators in adult rats.
\newblock {\em eLife}~5:\mbox{e14203}.

\bibitem[\protect\citeauthoryear{Huh  \bgroup et al.\egroup }{2016}]{huh2016}
Huh CYL, Amilhon B, Ferguson KA, Manseau F, Torres-Platas SG, Peach JP, Scodras
  S, Mechawar N, Skinner FK, Williams S (2016)
\newblock Excitatory {{Inputs Determine Phase}}-{{Locking Strength}} and
  {{Spike}}-{{Timing}} of {{CA1 Stratum Oriens}}/{{Alveus Parvalbumin}} and
  {{Somatostatin Interneurons}} during {{Intrinsically Generated Hippocampal
  Theta Rhythm}}.
\newblock {\em The Journal of Neuroscience: The Official Journal of the Society
  for Neuroscience}~36:\mbox{6605--6622}.

\bibitem[\protect\citeauthoryear{Jammalamadaka and
  SenGupta}{2001}]{jammalamadaka2001}
Jammalamadaka SR, SenGupta A (2001)
\newblock {\em Topics in {{Circular Statistics}}}
\newblock {World Scientific}.

\bibitem[\protect\citeauthoryear{Janczewski and Feldman}{2006}]{janczewski2006}
Janczewski WA, Feldman JL (2006)
\newblock Distinct rhythm generators for inspiration and expiration in the
  juvenile rat.
\newblock {\em The Journal of Physiology}~570:\mbox{407--420}.

\bibitem[\protect\citeauthoryear{Janczewski  \bgroup et al.\egroup
  }{2013}]{janczewski2013}
Janczewski WA, Tashima A, Hsu P, Cui Y, Feldman JL (2013)
\newblock Role of {{Inhibition}} in {{Respiratory Pattern Generation}}.
\newblock {\em The Journal of Neuroscience}~33:\mbox{5454--5465}.

\bibitem[\protect\citeauthoryear{Kam  \bgroup et al.\egroup }{2013}]{kam2013}
Kam K, Worrell JW, Janczewski WA, Cui Y, Feldman JL (2013)
\newblock Distinct {{Inspiratory Rhythm}} and {{Pattern Generating Mechanisms}}
  in the {{preBotzinger Complex}}.
\newblock {\em Journal of Neuroscience}~33:\mbox{9235--9245}.

\bibitem[\protect\citeauthoryear{Kiehn}{2011}]{kiehn2011}
Kiehn O (2011)
\newblock Development and functional organization of spinal locomotor circuits.
\newblock {\em Current Opinion in Neurobiology}~21:\mbox{100--109}.

\bibitem[\protect\citeauthoryear{Koizumi  \bgroup et al.\egroup
  }{2013}]{koizumi2013}
Koizumi H, Koshiya N, Chia JX, Cao F, Nugent J, Zhang R, Smith JC (2013)
\newblock Structural-{{Functional Properties}} of {{Identified Excitatory}} and
  {{Inhibitory Interneurons}} within {{Pre}}-{{B{\"o}tzinger Complex
  Respiratory Microcircuits}}.
\newblock {\em The Journal of Neuroscience}~33:\mbox{2994--3009}.

\bibitem[\protect\citeauthoryear{Koizumi and Smith}{2008}]{koizumi2008}
Koizumi H, Smith JC (2008)
\newblock Persistent {{Na}}+ and {{K}}+-{{Dominated Leak Currents Contribute}}
  to {{Respiratory Rhythm Generation}} in the {{Pre}}-{{B{\"o}tzinger Complex
  In Vitro}}.
\newblock {\em The Journal of Neuroscience}~28:\mbox{1773--1785}.

\bibitem[\protect\citeauthoryear{Kopell  \bgroup et al.\egroup
  }{2010}]{kopell2010}
Kopell N, Kramer MA, Malerba P, Whittington MA (2010)
\newblock Are {{Different Rhythms Good}} for {{Different Functions}}?
\newblock {\em Frontiers in Human Neuroscience}~4.

\bibitem[\protect\citeauthoryear{Lal  \bgroup et al.\egroup }{2011}]{lal2011}
Lal A, Oku Y, H{\"u}lsmann S, Okada Y, Miwakeichi F, Kawai S, Tamura Y,
  Ishiguro M (2011)
\newblock Dual oscillator model of the respiratory neuronal network generating
  quantal slowing of respiratory rhythm.
\newblock {\em Journal of Computational Neuroscience}~30:\mbox{225--240}.

\bibitem[\protect\citeauthoryear{Lewicki}{1998}]{lewicki1998}
Lewicki MS (1998)
\newblock A review of methods for spike sorting: The detection and
  classification of neural action potentials.
\newblock {\em Network (Bristol, England)}~9:\mbox{R53--78}.

\bibitem[\protect\citeauthoryear{Lieske  \bgroup et al.\egroup
  }{2000}]{lieske2000}
Lieske SP, Thoby-Brisson M, Telgkamp P, Ramirez JM (2000)
\newblock Reconfiguration of the neural network controlling multiple breathing
  patterns: Eupnea, sighs and gasps.
\newblock {\em Nature neuroscience}~3:\mbox{600--607}.

\bibitem[\protect\citeauthoryear{Lieske and Ramirez}{2006}]{lieske2006}
Lieske SP, Ramirez JM (2006)
\newblock Pattern-{{Specific Synaptic Mechanisms}} in a {{Multifunctional
  Network}}. {{I}}. {{Effects}} of {{Alterations}} in {{Synapse Strength}}.
\newblock {\em Journal of Neurophysiology}~95:\mbox{1323--1333}.

\bibitem[\protect\citeauthoryear{Lindsey \bgroup et al.\egroup
  }{2012}]{lindsey2012}
Lindsey BG, Rybak IA, Smith JC (2012)
\newblock Computational {{Models}} and {{Emergent Properties}} of {{Respiratory
  Neural Networks}}.
\newblock {\em Comprehensive Physiology}~2:\mbox{1619--1670}.

\bibitem[\protect\citeauthoryear{Marder and Bucher}{2001}]{marder2001}
Marder E, Bucher D (2001)
\newblock Central pattern generators and the control of rhythmic movements.
\newblock {\em Current Biology}~11:\mbox{R986--R996}.

\bibitem[\protect\citeauthoryear{Masuda and Aihara}{2004}]{masuda2004}
Masuda N, Aihara K (2004)
\newblock Global and local synchrony of coupled neurons in small-world
  networks.
\newblock {\em Biological Cybernetics}~90:\mbox{302--309}.

\bibitem[\protect\citeauthoryear{Missaghi  \bgroup et al.\egroup
  }{2016}]{missaghi2016}
Missaghi K, Le~Gal JP, Gray PA, Dubuc R (2016)
\newblock The neural control of respiration in lampreys.
\newblock {\em Respiratory Physiology \& Neurobiology}~234:\mbox{14--25}.

\bibitem[\protect\citeauthoryear{Molkov  \bgroup et al.\egroup
  }{2013}]{molkov2013}
Molkov YI, Bacak BJ, Dick TE, Rybak IA (2013)
\newblock Control of breathing by interacting pontine and pulmonary feedback
  loops.
\newblock {\em Frontiers in Neural Circuits}~7:\mbox{16}.

\bibitem[\protect\citeauthoryear{Moore  \bgroup et al.\egroup
  }{2013}]{moore2013}
Moore JD, Desch{\^e}nes M, Furuta T, Huber D, Smear MC, Demers M, Kleinfeld D
  (2013)
\newblock Hierarchy of orofacial rhythms revealed through whisking and
  breathing.
\newblock {\em Nature}~497:\mbox{205--210}.

\bibitem[\protect\citeauthoryear{Morgado-Valle \bgroup et al.\egroup
  }{2010}]{morgado-valle2010}
Morgado-Valle C, Baca SM, Feldman JL (2010)
\newblock Glycinergic {{Pacemaker Neurons}} in {{PreB{\"o}tzinger Complex}} of
  {{Neonatal Mouse}}.
\newblock {\em Journal of Neuroscience}~30:\mbox{3634--3639}.

\bibitem[\protect\citeauthoryear{Nieto-Posadas  \bgroup et al.\egroup
  }{2014}]{nieto-posadas2014}
Nieto-Posadas A, Flores-Mart{\~A}$\-$nez E,
  Lorea-Hern{\~A}\textexclamdown{}ndez JJ, Rivera-Angulo AJ,
  P{\~A}\textcopyright{}rez-Ortega JE, Bargas J, Pe{\~A}$\pm$a-Ortega F (2014)
\newblock Change in network connectivity during fictive-gasping generation in
  hypoxia: Prevention by a metabolic intermediate.
\newblock {\em Frontiers in Physiology}~5.

\bibitem[\protect\citeauthoryear{Onimaru \bgroup et al.\egroup
  }{2009}]{onimaru2009}
Onimaru H, Ikeda K, Kawakami K (2009)
\newblock Phox2b, {{RTN}}/{{pFRG}} neurons and respiratory rhythmogenesis.
\newblock {\em Respiratory Physiology \& Neurobiology}~168:\mbox{13--18}.

\bibitem[\protect\citeauthoryear{Onimaru  \bgroup et al.\egroup
  }{2015}]{onimaru2015}
Onimaru H, Tsuzawa K, Nakazono Y, Janczewski WA (2015)
\newblock Midline section of the medulla abolishes inspiratory activity and
  desynchronizes pre-inspiratory neuron rhythm on both sides of the medulla in
  newborn rats.
\newblock {\em Journal of Neurophysiology}~113:\mbox{2871--2878}.

\bibitem[\protect\citeauthoryear{Othman  \bgroup et al.\egroup
  }{2012}]{othman2012}
Othman NA, Gallacher M, Deeb TZ, Baptista-Hon DT, Perry DC, Hales TG (2012)
\newblock Influences on blockade by t-butylbicyclo-phosphoro-thionate of
  {{GABAA}} receptor spontaneous gating, agonist activation and
  desensitization.
\newblock {\em The Journal of Physiology}~590:\mbox{163--178}.

\bibitem[\protect\citeauthoryear{Park and Rubin}{2013}]{park2013}
Park C, Rubin JE (2013)
\newblock Cooperation of intrinsic bursting and calcium oscillations underlying
  activity patterns of model pre-{{B{\"o}tzinger}} complex neurons.
\newblock {\em Journal of Computational Neuroscience}~34:\mbox{345--366}.

\bibitem[\protect\citeauthoryear{Pe{\~n}a  \bgroup et al.\egroup
  }{2004}]{pena2004}
Pe{\~n}a F, Parkis MA, Tryba AK, Ramirez JM (2004)
\newblock Differential contribution of pacemaker properties to the generation
  of respiratory rhythms during normoxia and hypoxia.
\newblock {\em Neuron}~43:\mbox{105--117}.

\bibitem[\protect\citeauthoryear{Purvis  \bgroup et al.\egroup
  }{2006}]{purvis2006}
Purvis LK, Smith JC, Koizumi H, Butera RJ (2006)
\newblock Intrinsic {{Bursters Increase}} the {{Robustness}} of {{Rhythm
  Generation}} in an {{Excitatory Network}}.
\newblock {\em Journal of Neurophysiology}~97:\mbox{1515--1526}.

\bibitem[\protect\citeauthoryear{Ramirez \bgroup et al.\egroup
  }{1997}]{ramirez1997}
Ramirez JM, Quellmalz UJA, Wilken B (1997)
\newblock Developmental {{Changes}} in the {{Hypoxic Response}} of the
  {{Hypoglossus Respiratory Motor Output In Vitro}}.
\newblock {\em Journal of Neurophysiology}~78:\mbox{383--392}.

\bibitem[\protect\citeauthoryear{Ramirez  \bgroup et al.\egroup
  }{1998}]{ramirez1998}
Ramirez JM, Schwarzacher SW, Pierrefiche O, Olivera BM, Richter DW (1998)
\newblock Selective lesioning of the cat pre-{{B{\"o}tzinger}} complex in vivo
  eliminates breathing but not gasping.
\newblock {\em The Journal of Physiology}~507:\mbox{895--907}.

\bibitem[\protect\citeauthoryear{Ramirez  \bgroup et al.\egroup
  }{1997}]{ramirez1997a}
Ramirez JM, Telgkamp P, Elsen FP, Quellmalz UJ, Richter DW (1997)
\newblock Respiratory rhythm generation in mammals: Synaptic and membrane
  properties.
\newblock {\em Respiration Physiology}~110:\mbox{71--85}.

\bibitem[\protect\citeauthoryear{Ramirez  \bgroup et al.\egroup
  }{2016}]{ramirez2016}
Ramirez JM, Dashevskiy T, Marlin IA, Baertsch N (2016)
\newblock Microcircuits in respiratory rhythm generation: Commonalities with
  other rhythm generating networks and evolutionary perspectives.
\newblock {\em Current Opinion in Neurobiology}~41:\mbox{53--61}.

\bibitem[\protect\citeauthoryear{Ramirez  \bgroup et al.\egroup
  }{2011}]{ramirez2011}
Ramirez JM, Koch H, Garcia AJ, Doi A, Zanella S (2011)
\newblock The role of spiking and bursting pacemakers in the neuronal control
  of breathing.
\newblock {\em Journal of Biological Physics}~37:\mbox{241--261}.

\bibitem[\protect\citeauthoryear{Ramirez and Lieske}{2003}]{ramirez2003}
Ramirez JM, Lieske SP (2003)
\newblock Commentary on the definition of eupnea and gasping.
\newblock {\em Respiratory Physiology \& Neurobiology}~139:\mbox{113--119}.

\bibitem[\protect\citeauthoryear{Rekling \bgroup et al.\egroup
  }{2000}]{rekling2000}
Rekling JC, Shao XM, Feldman JL (2000)
\newblock Electrical {{Coupling}} and {{Excitatory Synaptic Transmission}}
  between {{Rhythmogenic Respiratory Neurons}} in the {{PreB{\"o}tzinger
  Complex}}.
\newblock {\em The Journal of Neuroscience}~20:\mbox{RC113--RC113}.

\bibitem[\protect\citeauthoryear{Richter and Smith}{2014}]{richter2014}
Richter DW, Smith JC (2014)
\newblock Respiratory {{Rhythm Generation In Vivo}}.
\newblock {\em Physiology}~29:\mbox{58--71}.

\bibitem[\protect\citeauthoryear{Rubin  \bgroup et al.\egroup
  }{2009}]{rubin2009}
Rubin JE, Shevtsova NA, Ermentrout GB, Smith JC, Rybak IA (2009)
\newblock Multiple {{Rhythmic States}} in a {{Model}} of the {{Respiratory
  Central Pattern Generator}}.
\newblock {\em Journal of Neurophysiology}~101:\mbox{2146--2165}.

\bibitem[\protect\citeauthoryear{Rubin  \bgroup et al.\egroup
  }{2011}]{rubin2011}
Rubin JE, Bacak BJ, Molkov YI, Shevtsova NA, Smith JC, Rybak IA (2011)
\newblock Interacting oscillations in neural control of breathing: Modeling and
  qualitative analysis.
\newblock {\em Journal of Computational Neuroscience}~30:\mbox{607--632}.

\bibitem[\protect\citeauthoryear{Rubin  \bgroup et al.\egroup
  }{2009}]{rubin2009a}
Rubin JE, Hayes JA, Mendenhall JL, Del~Negro CA (2009)
\newblock Calcium-activated nonspecific cation current and synaptic depression
  promote network-dependent burst oscillations.
\newblock {\em Proceedings of the National Academy of
  Sciences}~106:\mbox{2939--2944}.

\bibitem[\protect\citeauthoryear{Sara}{2009}]{sara2009}
Sara SJ (2009)
\newblock The locus coeruleus and noradrenergic modulation of cognition.
\newblock {\em Nature Reviews Neuroscience}~10:\mbox{211--223}.

\bibitem[\protect\citeauthoryear{Schwab  \bgroup et al.\egroup
  }{2010}]{schwab2010}
Schwab DJ, Bruinsma RF, Feldman JL, Levine AJ (2010)
\newblock Rhythmogenic neuronal networks, emergent leaders, and k-cores.
\newblock {\em Physical Review E}~82.

\bibitem[\protect\citeauthoryear{Schwarzacher \bgroup et al.\egroup
  }{2011}]{schwarzacher2011}
Schwarzacher SW, R{\"u}b U, Deller T (2011)
\newblock Neuroanatomical characteristics of the human pre-{{B{\"o}tzinger}}
  complex and its involvement in neurodegenerative brainstem diseases.
\newblock {\em Brain}~134:\mbox{24--35}.

\bibitem[\protect\citeauthoryear{Shao and Feldman}{1997}]{shao1997}
Shao XM, Feldman JL (1997)
\newblock Respiratory rhythm generation and synaptic inhibition of expiratory
  neurons in pre-{{B{\"o}tzinger}} complex: Differential roles of glycinergic
  and {{GABAergic}} neural transmission.
\newblock {\em Journal of Neurophysiology}~77:\mbox{1853--1860}.

\bibitem[\protect\citeauthoryear{Sharp \bgroup et al.\egroup
  }{1996}]{sharp1996}
Sharp AA, Skinner FK, Marder E (1996)
\newblock Mechanisms of oscillation in dynamic clamp constructed two-cell
  half-center circuits.
\newblock {\em Journal of Neurophysiology}~76:\mbox{867--883}.

\bibitem[\protect\citeauthoryear{Sherman  \bgroup et al.\egroup
  }{2015}]{sherman2015}
Sherman D, Worrell JW, Cui Y, Feldman JL (2015)
\newblock Optogenetic perturbation of {{preBotzinger}} complex inhibitory
  neurons modulates respiratory pattern.
\newblock {\em Nature Neuroscience}~18:\mbox{408--414}.

\bibitem[\protect\citeauthoryear{Skinner}{2012}]{skinner2012}
Skinner FK (2012)
\newblock Cellular-based modeling of oscillatory dynamics in brain networks.
\newblock {\em Current Opinion in Neurobiology}~22:\mbox{660--669}.

\bibitem[\protect\citeauthoryear{Smith  \bgroup et al.\egroup
  }{1991}]{smith1991}
Smith JC, Ellenberger HH, Ballanyi K, Richter DW, Feldman JL (1991)
\newblock Pre-{{Botzinger}} complex: A brainstem region that may generate
  respiratory rhythm in mammals.
\newblock {\em Science}~254:\mbox{726--729}.

\bibitem[\protect\citeauthoryear{Smith  \bgroup et al.\egroup
  }{2007}]{smith2007}
Smith JC, Abdala APL, Koizumi H, Rybak IA, Paton JF (2007)
\newblock Spatial and functional architecture of the mammalian brain stem
  respiratory network: A hierarchy of three oscillatory mechanisms.
\newblock {\em Journal of neurophysiology}~98:\mbox{3370--3387}.

\bibitem[\protect\citeauthoryear{Smith  \bgroup et al.\egroup
  }{2013}]{smith2013}
Smith JC, Abdala AP, Borgmann A, Rybak IA, Paton JF (2013)
\newblock Brainstem respiratory networks: Building blocks and microcircuits.
\newblock {\em Trends in Neurosciences}~36:\mbox{152--162}.

\bibitem[\protect\citeauthoryear{Stuart and Hultborn}{2008}]{stuart2008}
Stuart DG, Hultborn H (2008)
\newblock Thomas {{Graham Brown}} (1882\textendash{}1965), {{Anders Lundberg}}
  (1920\textendash), and the neural control of stepping.
\newblock {\em Brain Research Reviews}~59:\mbox{74--95}.

\bibitem[\protect\citeauthoryear{Tan  \bgroup et al.\egroup }{2008}]{tan2008}
Tan W, Janczewski WA, Yang P, Shao XM, Callaway EM, Feldman JL (2008)
\newblock Silencing {{preB{\"o}tzinger Complex}} somatostatin-expressing
  neurons induces persistent apnea in awake rat.
\newblock {\em Nature Neuroscience}~11:\mbox{538--540}.

\bibitem[\protect\citeauthoryear{Tan  \bgroup et al.\egroup }{2010}]{tan2010}
Tan W, Pagliardini S, Yang P, Janczewski WA, Feldman JL (2010)
\newblock Projections of {{preB{\"o}tzinger Complex}} neurons in adult rats.
\newblock {\em The Journal of Comparative Neurology}~518:\mbox{1862--1878}.

\bibitem[\protect\citeauthoryear{Thoby-Brisson and
  Ramirez}{2001}]{thoby-brisson2001}
Thoby-Brisson M, Ramirez JM (2001)
\newblock Identification of {{Two Types}} of {{Inspiratory Pacemaker Neurons}}
  in the {{Isolated Respiratory Neural Network}} of {{Mice}}.
\newblock {\em Journal of Neurophysiology}~86:\mbox{104--112}.

\bibitem[\protect\citeauthoryear{Toporikova and Butera}{2011}]{toporikova2011}
Toporikova N, Butera RJ (2011)
\newblock Two types of independent bursting mechanisms in inspiratory neurons:
  An integrative model.
\newblock {\em Journal of Computational Neuroscience}~30:\mbox{515--528}.

\bibitem[\protect\citeauthoryear{Wang  \bgroup et al.\egroup }{2014}]{wang2014}
Wang X, Hayes JA, Revill AL, Song H, Kottick A, Vann NC, LaMar MD, Picardo MCD,
  Akins VT, Funk GD, {others} (2014)
\newblock Laser ablation of {{Dbx1}} neurons in the pre-{{B{\"o}tzinger}}
  complex stops inspiratory rhythm and impairs output in neonatal mice.
\newblock {\em eLife}~3:\mbox{e03427}.

\bibitem[\protect\citeauthoryear{Winter  \bgroup et al.\egroup
  }{2009}]{winter2009}
Winter SM, Fresemann J, Schnell C, Oku Y, Hirrlinger J, H{\"u}lsmann S (2009)
\newblock Glycinergic interneurons are functionally integrated into the
  inspiratory network of mouse medullary slices.
\newblock {\em Pfl{\"u}gers Archiv - European Journal of
  Physiology}~458:\mbox{459--469}.

\bibitem[\protect\citeauthoryear{Wittmeier  \bgroup et al.\egroup
  }{2008}]{wittmeier2008}
Wittmeier S, Song G, Duffin J, Poon CS (2008)
\newblock Pacemakers handshake synchronization mechanism of mammalian
  respiratory rhythmogenesis.
\newblock {\em Proceedings of the National Academy of
  Sciences}~105:\mbox{18000--18005}.

\end{thebibliography}

\end{document}